\documentclass{article}
\usepackage{graphicx}
\usepackage{cite}

\begin{document}

\title{Information-Theoretic Signatures of Biodiversity in the Barcoding Gene}

\author{Valmir~C.~Barbosa\thanks{valmir@cos.ufrj.br}\\
\\
Programa de Engenharia de Sistemas e Computa\c c\~ao, COPPE\\
Universidade Federal do Rio de Janeiro\\
Caixa Postal 68511\\
21941-972 Rio de Janeiro - RJ, Brazil}

\date{}

\maketitle

\begin{abstract}
The COI mitochondrial gene is present in all animal phyla and in a few others,
and is the leading candidate for species identification through DNA barcoding.
Calculating a generalized form of total correlation on publicly available data
on the gene yields distinctive information-theoretic descriptors of the phyla
represented in the data. Moreover, performing principal component analysis on
standardized versions of these descriptors reveals a strong correlation between
the first principal component and the natural logarithm of the number of known
living species. The descriptors thus constitute clear information-theoretic
signatures of the processes whereby evolution has given rise to current
biodiversity.

\end{abstract}

\newpage
\section{Introduction}

The genome of every living organism is the product of Darwinian natural
selection and random drift as they played out along the ages since the
inception of life on the planet. The fundamental hallmark of this lengthy
process has been the appearance and diversification of highly organized (out of
highly unorganized) matter. It is therefore only expected that any pool of DNA
from a sufficiently representative set of individuals should contain
unmistakable information-theoretic traces of how the reduction of entropy that
the evolution of life has entailed led to the biodiversity that we observe
today. This notwithstanding, previous efforts to analyze the information content
of DNA have been limited in scope and as such provided little or no insight into
biodiversity \cite{g66,c97,sh97,sb16}.

As we see it, there are two main difficulties to be faced when attempting such
analyses. The first one has to do with the availability of experimental data, as
well as with the quality of such data when they are available. Any
information-theoretic analysis of DNA has to begin by grappling with estimating
the distribution of probability associated with the pool that is being studied.
If the pool comes, say, from a single genus, and given some individual of that
genus, we must be able to estimate the probability that the genus contains DNA
that is sufficiently similar to that observed for the given individual.
Moreover, this only makes sense if the DNA in question comes from a region of
the genome that is found in all individuals in that genus, which further
complicates the data availability and quality problem. The second difficulty has
to do with formulating the problem adequately. While the well-known Shannon
entropy can be effectively used to quantify information gain, it needs to be
applied carefully in order to tease out the desired footprints of biodiversity.

Here we address the first difficulty by taking advantage of the current public
availability of data on the so-called barcoding gene (the COI mitochondrial
gene) from the BOLD (Barcode Of Life Data) systems initiative \cite{rh07}. This
gene is present in all animals as well as in members of a few other phyla
\cite{hcbd03}. The BOLD repository contains a few million sequence fragments of
the gene's $710$-base-pair Folmer region \cite{fbhlv94}, covering $25$ phyla, of
which $23$ are animal phyla and the remaining two are phyla of algae. The
barcoding gene is therefore ubiquitous to a large extent. It is also diverse
enough across species (thence its denomination as a ``barcode'' \cite{scvrl05}),
so the BOLD samples are poised to constitute a suitable dataset from which to
estimate the necessary probability distributions. Owing to the way in which
these data are organized, we perform our study at the level of the phylum.

As for the difficulty of formulating the problem adequately, we follow the
principle first expressed qualitatively by Rothstein in 1952 regarding the
analysis of information gain in the multivariate case \cite{r52}. This principle
considers a set of random variables representing some physical process of
interest, as well as their joint distribution. It states that, since information
gain occurs during the process not only at the level of the set of all variables
taken as a whole but also at the level of any of its subsets, what really
matters is the amount of global gain that surpasses the combined local gains
relative to select subsets. This principle has been formalized as the so-called
total correlation of the variables and its generalizations \cite{w60}. It is
behind several studies related to the integration of information in complex
systems, including some targeted at the temporal evolution of probabilistic
cellular automata \cite{cb15}, of threshold-based systems in general \cite{b17},
and of the cerebral cortex as it gives rise to consciousness
\cite{bt08,nb11,oat14}.

We use a particular form of generalized total correlation to obtain
information-theoretic descriptors of the barcoding gene for each of the phyla
considered. These descriptors are correlated with one another to various
degrees, but mapping them onto the uncorrelated, variance-emphasizing directions
provided by PCA (principal component analysis) \cite{aw10} reveals them to be
signatures of biodiversity that lie hidden in the gene. Specifically, for each
phylum a single new descriptor is obtained that correlates strongly with the
biodiversity that is present in the phylum as expressed by the logarithm of its
number of species (either those represented in the data or the many more that
are known but have not yet reached the BOLD system). Biodiversity, therefore, is
seen to grow exponentially with that single descriptor of a phylum.

\section{Generalized total correlation}

Let $S=\langle S_1,S_2,\ldots,S_n\rangle$ be an $n$-nucleotide sequence, i.e.,
each $S_k$ is one of the four nucleotides occurring in DNA (represented by base
A, C, G, or T), and let $\mathcal{B}_n$ be the set of all $4^n$ $n$-nucleotide
sequences. We assume that $n$ is a multiple of $3$ (the number of nucleotides in
a codon) and view each $S_k$ as a random variable. The form of generalized total
correlation that we use is based on partitioning $S$ into contiguous
subsequences. The number of possible partitions is exponential in $n$, which
probably would rule out any attempt at choosing an optimal one given the
ultimate goal of exposing biodiversity even if such optimality were well
defined. The partition we select requires every subsequence to be $d<n$
nucleotides long, with $n$ a multiple of $d$ and $d$ a multiple of $3$. The
number of subsequences in the partition is $n/d$.

Our version of generalized total correlation is specific to sequence $S$ of
random variables and divisor $d$. It is denoted by $C(S,d)$ and given by the
Kullback-Leibler divergence from the joint distribution of the $n$ random
variables that would ensue if all $n/d$ subsequences were probabilistically
independent of one another to the actual joint distribution. That is,
\begin{equation}
C(S,d)=
\sum_{T\in\mathcal{B}_n}\mathrm{Pr}(S=T)
\log_4
\frac
{\mathrm{Pr}(S=T)}
{\prod_{i=1}^{n/d}\mathrm{Pr}(S_{id-d+1}^{id}=T_{id-d+1}^{id})},
\end{equation}
where each $S_k^\ell$, with $\ell-k+1=d$, denotes the $d$-nucleotide sequence
beginning at nucleotide $S_k$ and ending at nucleotide $S_\ell$, and similarly
for $T_k^\ell$. This expression can be rewritten as
\begin{equation}
C(S,d)=\sum_{i=1}^{n/d}H(S_{id-d+1}^{id})-H(S),
\label{eq:gtc}
\end{equation}
where
\begin{equation}
H(S_k^\ell)=
-\sum_{T\in\mathcal{B}_d}
\mathrm{Pr}(S_k^\ell=T_k^\ell)\log_4\mathrm{Pr}(S_k^\ell=T_k^\ell)
\end{equation}
and
\begin{equation}
H(S)=
-\sum_{T\in\mathcal{B}_n}\mathrm{Pr}(S=T)\log_4\mathrm{Pr}(S=T)
\end{equation}
give the Shannon entropy of $S_k^\ell$ and that of $S$, respectively
($\mathcal{B}_d$ is the set of all $4^d$ $d$-nucleotide sequences). The use of
logarithms to the base $4$ implies $H(S_k^\ell)\le d$ for each pair $k,\ell$ and
$H(S)\le n$. These bounds, in turn, suggest a reinterpretation of
Eq.~(\ref{eq:gtc}), after rewriting it in the form
\begin{equation}
C(S,d)=[n-H(S)]-\sum_{i=1}^{n/d}[d-H(S_{id-d+1}^{id})],
\end{equation}
where each term in square brackets is an information gain, or reduction of
entropy, relative to the case of maximum entropy. The reinterpretation,
therefore, is of $C(S,d)$ as the amount of global information gain (i.e.,
relative to sequence $S$) that surpasses the sum total of the local information
gains (i.e., relative to the subsequences $S_k^\ell$). $C(S,d)$ is expressed in
quaternary digits (\emph{quats}), but henceforth we use its normalized version,
denoted by $N(S,d)$ and given by
\begin{equation}
N(S,d)=\frac{C(S,d)}{n-d},
\end{equation}
whose denominator gives the maximum possible value of $C(S,d)$.

\section{Methods}

All data publicly available at the BOLD website (barcodinglife.org) on September
20, 2017 were downloaded in the FASTA format. All sequences start at the
$5'$-end of the barcoding gene but not all have the same number of nucleotides.
Moreover, sequences often contain symbols other than A, C, G, or T, following
the FASTA convention that hyphens and other eleven letters can also appear and
have specific meanings: a hyphen denotes a gap of indeterminate length produced
during sequencing; the other letters indicate uncertainty in tagging the signal
that is read off the sequencer with one of the four base letters (for example, R
indicates uncertainty between A and G; B indicates uncertainty among the non-A
bases; and N indicates full uncertainty).

In order to allow each available sequence to contribute as fully as possible,
each one can give rise to several others, each of a different length. This is
achieved by truncating the original sequence to its first $432$ nucleotides,
then to the first $459$ nucleotides, then $486$, and so on through $702$
nucleotides. Depending on how each of these numbers relates to the original
length, each sequence can yield anywhere from none up to eleven sequences for
use in the study, possibly including one with the $648$ nucleotides of the
canonical barcoding gene fragment \cite{rh07}.

Before participating in the estimation of the joint distribution, each sequence
thus generated is screened and discarded if either of two flags is raised. The
first one indicates the presence of a hyphen. The second indicates the presence
of too many letters other than A, C, G, or T. In order to decide whether to
raise this second flag, we first assign an unfolding factor to each letter in
the sequence. The factor assigned to the $k$th letter is $q_k=1$ if the letter
is A, C, G, or T; $q_k=2$ if the letter expresses uncertainty between two of
those four (such as R, used above in the example); $q_k=3$ if the uncertainty is
among three of them (such as B in the example); or $q_k=4$ if the letter is N.
The flag is raised if $Q > 1\,000$, where $Q=\prod_{k=1}^n q_k$, which is meant
to ensure that the storage required for computing does not run out of bounds
during distribution estimation.

The number of surviving sequences for each phylum and for each of
$n=432,459,\ldots,702$ is shown in Table~\ref{nseqs}. This has led to the
elimination of two phyla from the study, Cycliophora and Gnathostomulida, due
to the absence, for at least one value of $n$, of surviving sequences. We
therefore proceed with a total of $23$ phyla. For each of these phyla, the total
number of $N(S,d)$ descriptors is $107$, all eleven values of $n$ considered,
every multiple-of-$3$ divisor $d$ of each $n$ considered as well (except for $n$
itself).

For each phylum and each value of $n$, the essential probability to be
estimated is the joint probability $\mathrm{Pr}(S=T)$ for each
$T\in\mathcal{B}_n$, since from these all marginals related to the subsequences
of $S$ can be computed. A first approach to such estimation, the so-called
plug-in approach, is to first count the number of occurrences of $T$ among the
sequences that survived screening and then normalize the counts. Because
$\mathcal{B}_n$ only contains sequences of the base letters A, C, G, and T, each
surviving sequence is first unfolded into $Q$ sequences that only have base
letters as well. Each of these $Q$ sequences includes one full set of the
possible replacements of the non-base letters with base letters that resolve the
uncertainties, and receives a contribution of $1/Q$ to its count.

This approach sets $\mathrm{Pr}(S=T)$ to $0$ for nearly all $T\in\mathcal{B}_n$,
which can sometimes be too imprecise due to the limited number of surviving
sequences (even after unfolding). There would be nothing else to be done if
distribution estimation were the final goal, but for entropy calculations from
the estimated distribution there are several approaches that seek to alleviate
the problem \cite{p03,hs09,app14}. In this study we use the Hausser-Strimmer
shrinkage approach \cite{hs09}, which essentially begins with the same counting
as above and then alters the resulting probabilities via a data-dependent convex
combination with the uniform distribution over $\mathcal{B}_n$.

A crucial methodological ingredient is to drastically simplify the description
of each phylum as a point in $107$ dimensions to one in only a few dimensions.
We do this by viewing each of the $107$ $N(S,d)$ descriptors as a random
variable for which $23$ samples are available (one for each phylum) and using
PCA on these variables' standardized versions (obtained by shifting and scaling
each variable's samples so that the resulting mean and standard deviation are
$0$ and $1$, respectively). PCA returns a description of the phyla in terms of
principal components, that is, uncorrelated projections of the random variables'
standardized samples. The first component accounts for more variance across the
phyla than the second, the second more than the third, and so on. One then
retains as many of these components as needed to account for as much variance as
desired, hopefully only a few of them. Projecting a phylum's $107$-dimensional
vector of standardized samples onto one of the retained components is achieved
by a zero-intercept linear regression through the $107$ coefficients that PCA
outputs, known as that component's loadings.

\section{Results}

\begin{figure}[p]
\centering
\includegraphics[scale=0.80]{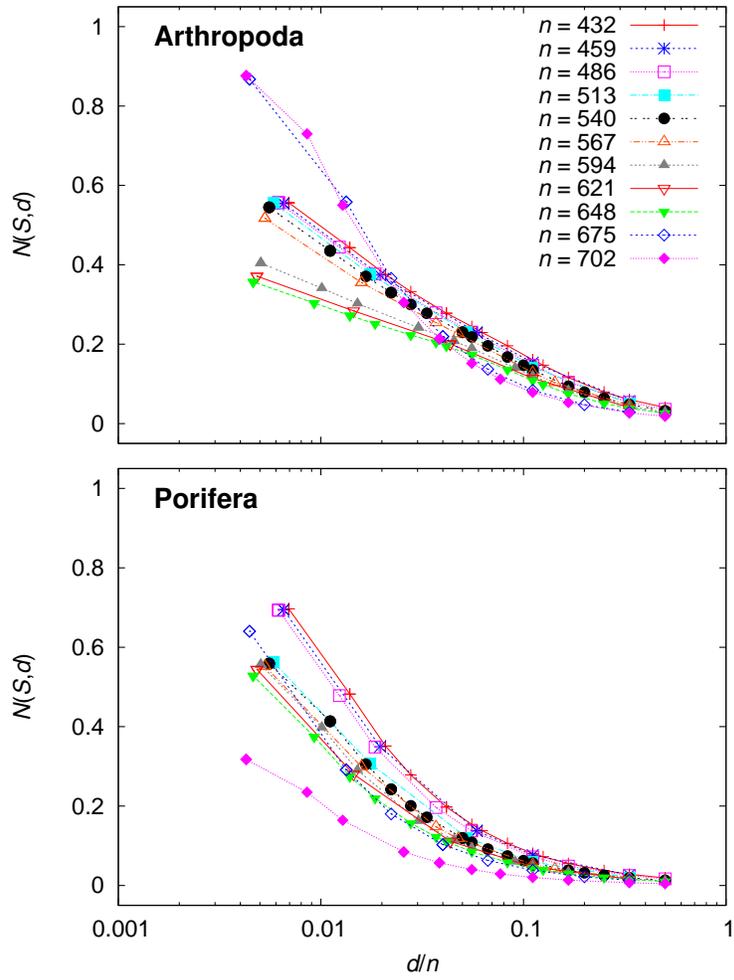}\\
\caption{The $107$ $N(S,d)$ descriptors for the phyla Arthropoda and Porifera.}
\label{ngtc}
\end{figure}

For each of the $23$ phyla whose BOLD data survived screening we obtained all
$107$ $N(S,d)$ descriptors as outlined. Drawing a scatter plot for each phylum
with each descriptor represented by its $d/n$ and $N(S,d)$ values yields a rich
variety of possibilities, as illustrated in Figure~\ref{ngtc-app} for all phyla
and in Figure~\ref{ngtc} for Arthropoda and Porifera. These two phyla were
singled out because they illustrate particularly well how striking the
difference between the descriptors of distinct phyla can be. In fact, even
though for any given phylum the value of $N(S,d)$ decreases with increasing $d$
for fixed $n$ [note that the unnormalized value of $N(S,d)$, $C(S,d)$, equals
$0$ for $d=n$, regardless of the underlying joint distribution], visually it is
clear that sometimes the descriptors of two given phyla share not much else.
 
\begin{figure}[t]
\centering
\includegraphics[scale=0.80]{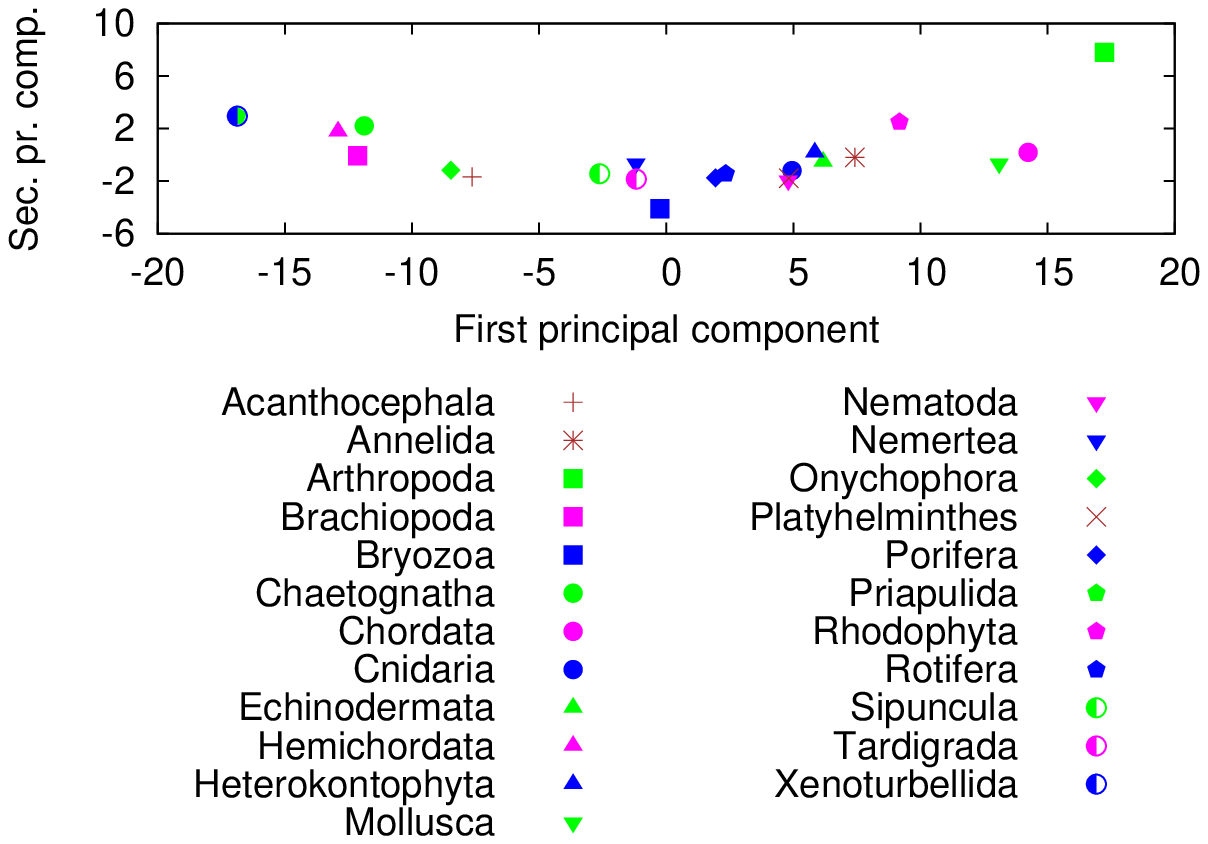}\\
\caption{Each phylum as represented by the first and second principal components
of all $107$ standardized $N(S,d)$ descriptors. The symbol for Priapulida is
covered by that for Xenoturbellida.}
\label{PCs}
\end{figure}

Applying PCA yields a first principal component that accounts for $90.36\%$ of
the variance across the phyla, a second component accounting for $5.67\%$ of
such variance, and so on, as shown in the so-called scree plot of
Figure~\ref{scree}. Projecting the $107$-dimensional standardized phylum
samples onto the first two principal components yields the scatter plot shown in
Figure~\ref{PCs}, where therefore over $96\%$ of the variance is accounted for.
The first principal component, in particular, suggests a nearly unique ordering
of the phyla, from the ones with the most negative values (Priapulida and
Xenoturbellida) to the one with the most positive value (Arthropoda). This order
is used in Table~\ref{nspecs} to list all phyla along with two numbers of
species for each phylum. The fist one (N.Sp.) is the number of species accounted
for in the data for that phylum and the second (N.Liv.Sp.) is the number of
known living species the phylum has. N.Sp.\ is a lower bound on the number of
species represented in the data, since not all sequences are tagged with a
species name. Note that, regardless of which species count one focuses on, there
is a general tendency for the number of species to grow rapidly as the phyla are
considered in the order of increasing first principal component.
 
\begin{figure}[p]
\centering
\includegraphics[scale=0.80]{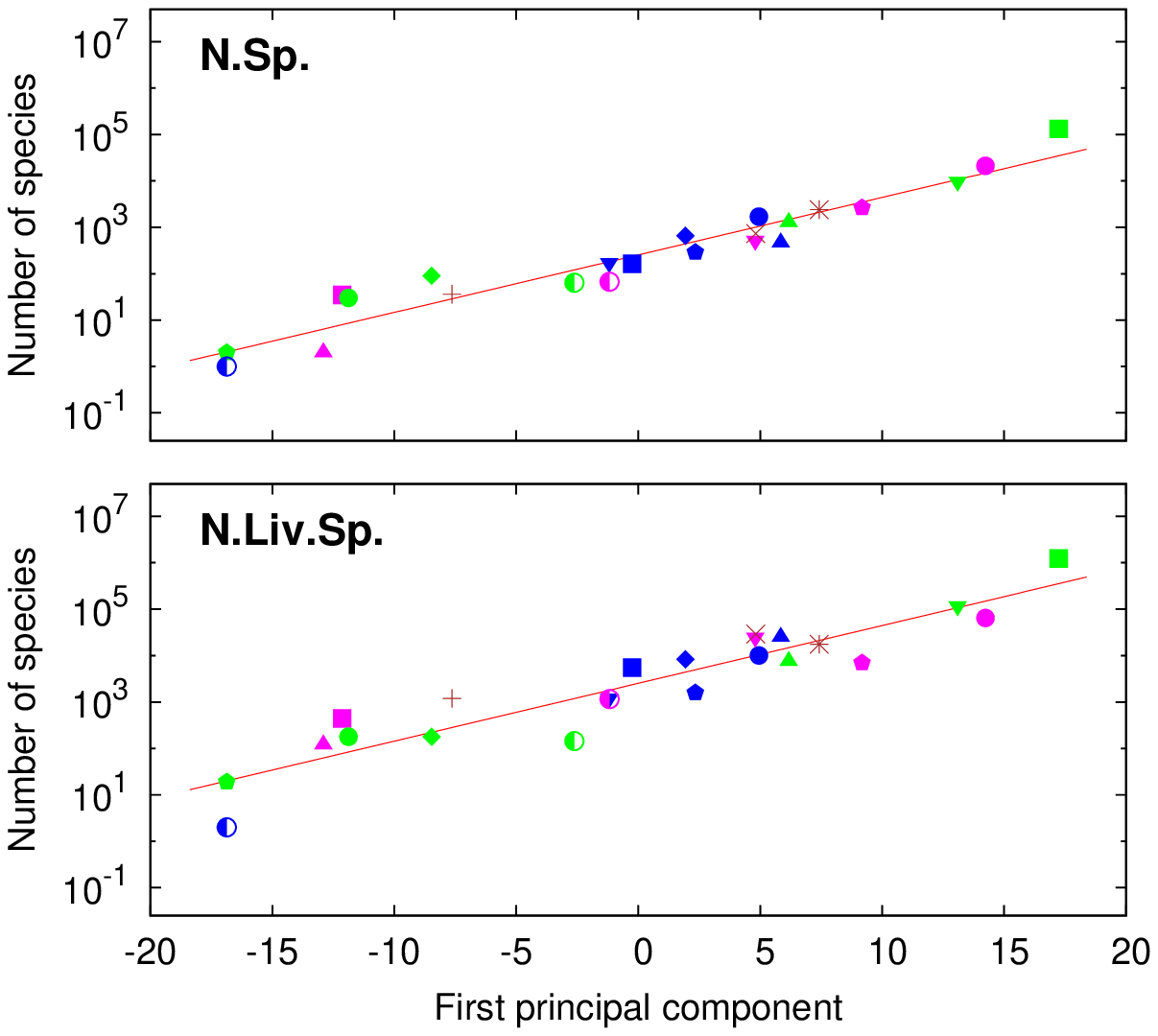}\\
\caption{Each phylum as represented by the first principal component of all
$107$ standardized $N(S,d)$ descriptors and by the number of species in the
phylum (either the lower bound given by the data, N.Sp., or the number of known
living species, N.Liv.Sp.; see Table~\ref{nspecs}). The fitted linear models for
N.Sp.\ and N.Liv.Sp.\ are
$\ln y=5.53792+0.28535x$ (p-value: $2.256\times 10^{-12}$)
and
$\ln y=7.83349+0.28680x$ (p-value: $7.982\times 10^{-11}$),
respectively. Symbol keys are the same as in Figure~\ref{PCs}.}
\label{lm}
\end{figure}

In fact, computing the Pearson correlation coefficient between the first
principal component and the natural logarithm of the number of species over all
$23$ phyla yields a little over $0.96$ for N.Sp., a little over $0.93$ for
N.Liv.Sp. The corresponding linear models are given in Figure~\ref{lm}. We find
it a striking feature of these models that, in both cases, their slopes are
practically the same: $0.28535$ and $0.28680$, respectively for N.Sp.\ and
N.Liv.Sp., the latter only $0.5\%$ above the former. Thus, not only are the
$N(S,d)$ descriptors strong information-theoretic signatures of a phylum's
biodiversity as annotated in the available data on the barcoding gene, they are
also seen to be robust signatures, since they relate nearly as significantly to
all the known biodiversity that is mostly absent from the data.

\begin{table}[p]
\centering
\caption{Lower bound on the number of species represented in the data (N.Sp.),
as reported on the BOLD website (barcodinglife.org) on September 20, 2017; and
number of known living species (N.Liv.Sp.), as reported on the Encyclopedia of
Life website (eol.org/collections/18879) on November 1, 2017 (two exceptions are
the N.Liv.Sp.\ values for Heterokontophyta and Rhodophyta, obtained from
wikipedia.org/wiki/Heterokont and wikipedia.org/wiki/Red\_algae, respectively,
on October 21, 2017). Phyla are listed in the order of increasing first
principal component, as implied by Figure~\ref{PCs}. A tie exists between
Priapulida and Xenoturbellida, which are then listed in the order of increasing
N.Sp. (or N.Liv.Sp.).}
\label{nspecs}
\begin{tabular}{lrr}
\hline
\multicolumn{1}{c}{Phylum}
&\multicolumn{1}{c}{N.Sp.}
&\multicolumn{1}{c}{N.Liv.Sp.} \\
\hline
Xenoturbellida   &        $1$ &           $2$ \\
Priapulida       &        $2$ &          $19$ \\
Hemichordata     &        $2$ &         $120$ \\
Brachiopoda      &       $35$ &         $443$ \\
Chaetognatha     &       $30$ &         $179$ \\
Onychophora      &       $90$ &         $179$ \\
Acanthocephala   &       $36$ &      $1\,192$ \\
Sipuncula        &       $64$ &         $144$ \\
Nemertea         &      $172$ &      $1\,200$ \\
Tardigrada       &       $67$ &      $1\,157$ \\
Bryozoa          &      $163$ &      $5\,486$ \\
Porifera         &      $656$ &      $8\,346$ \\
Rotifera         &      $289$ &      $1\,583$ \\
Nematoda         &      $521$ &     $24\,773$ \\
Platyhelminthes  &      $729$ &     $29\,285$ \\
Cnidaria         &   $1\,702$ &     $10\,105$ \\
Heterokontophyta &      $465$ &     $25\,000$ \\
Echinodermata    &   $1\,278$ &      $7\,509$ \\
Annelida         &   $2\,400$ &     $17\,446$ \\
Rhodophyta       &   $2\,659$ &      $7\,000$ \\
Mollusca         &   $9\,768$ &    $117\,358$ \\
Chordata         &  $21\,035$ &     $64\,791$ \\
Arthropoda       & $132\,834$ & $1\,235\,858$ \\
\hline
\end{tabular}

\end{table}

\section{Conclusion and outlook}

\begin{figure}[t]
\centering
\includegraphics[scale=0.80]{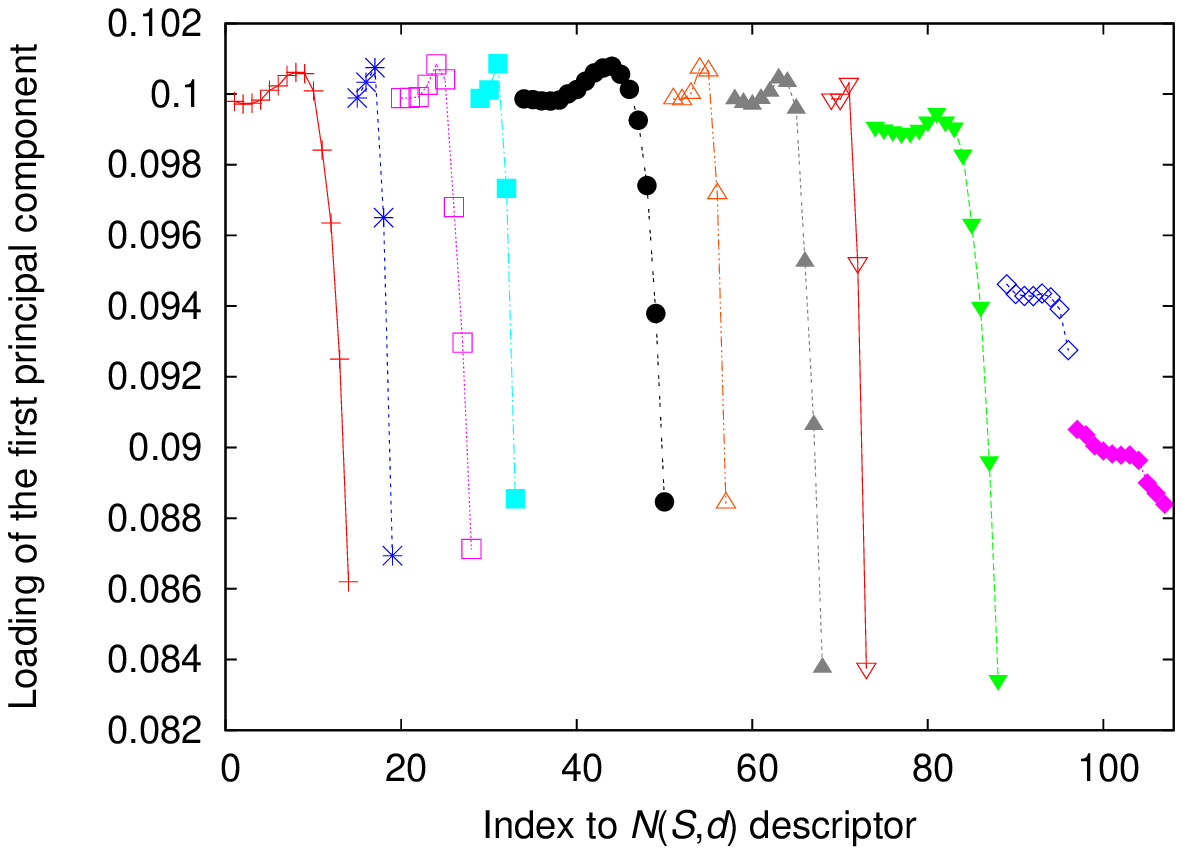}\\
\caption{Loadings of the first principal component for each of the $107$
standardized $N(S,d)$ descriptors. Descriptors are indexed upward from $1$,
arranged lexicographically on the pair $n,d$. Symbol keys are the same as in
Figure~\ref{ngtc}.}
\label{loadings}
\end{figure}

The discovery related in this paper has been driven mainly by data, following aninitial intuition regarding the nature of information gain as expressed by the
generalized forms of total correlation, as well as the primacy of the codon,
rather than the nucleotide, as the most meaningful structural unit. While this
data-centric approach may seem only natural in an age of successful machine
learning and data science in general, we believe such discoveries beg many more
questions than they answer. In the case at hand, a principled theory supporting
what has been discovered is still lacking and should be pursued. One necessary
first step toward further understanding is to interpret the loadings of the
first principal component. These are shown in Figure~\ref{loadings}, where we
see that loadings are less influential for the largest values of $n$. Likewise,
for $n\le 648$ they peak at around $d=27$--$54$ nucleotides, that is, $9$--$18$
codons.

Moreover, our claim that the $N(S,d)$ descriptors constitute robust signatures
of biodiversity (since they lead to essentially same-slope linear models
regardless of whether the number of species in question is that taken from the
data or that of the known living species) should stand in the face of further
scrutiny. In this regard, not only does the BOLD repository tend to grow
continually, but perhaps more importantly, most biodiversity and conservation
specialists seem convinced that the number of undiscovered living species
surpasses that of known living species by a wide margin
\cite{mtasw11,sjpl12,lmrw17}. Robustness is then to be reevaluated whenever the
availability of data or the number of known living species gets substantially
modified.

\subsection*{Acknowledgments}

We acknowledge partial support from CNPq, CAPES, and a FAPERJ BBP grant.

\bibliography{barcoding}

\begin{thebibliography}{10}

\bibitem{g66}
L.~L. Gatlin.
\newblock The information content of {DNA}.
\newblock {\em J. Theor. Biol.}, 10:281--300, 1966.

\bibitem{c97}
R.~H. Crozier.
\newblock Preserving the information content of species: genetic diversity,
  phylogeny, and conservation worth.
\newblock {\em Annu. Rev. Ecol. Syst.}, 28:243--268, 1997.

\bibitem{sh97}
A.~O. Schmitt and H.~Herzel.
\newblock Estimating the entropy of {DNA} sequences.
\newblock {\em J. Theor. Biol.}, 188:369--377, 1997.

\bibitem{sb16}
S.~Srivastava and M.~S. Baptista.
\newblock Markovian language model of the {DNA} and its information content.
\newblock {\em Roy. Soc. Open Sci.}, 3:150527, 2016.

\bibitem{rh07}
S.~Ratnasingham and P.~D.~N. Hebert.
\newblock {BOLD}: the barcode of life data system (www.barcodinglife.org).
\newblock {\em Mol. Ecol. Notes}, 7:355--364, 2007.

\bibitem{hcbd03}
P.~D.~N. Hebert, A.~Cywinska, S.~L. Ball, and J.~R. {deWaard}.
\newblock Biological identifications through {DNA} barcodes.
\newblock {\em Proc. R. Soc. Lond. B}, 270:313--321, 2003.

\bibitem{fbhlv94}
O.~Folmer, M.~Black, W.~Hoeh, R.~Lutz, and R.~Vrijenhoek.
\newblock {DNA} primers for amplification of mitochondrial cytochrome \emph{c}
  oxidase subunit {I} from diverse metazoan invertebrates.
\newblock {\em Mol. Mar. Biol. Biotech.}, 3:294--299, 1994.

\bibitem{scvrl05}
V.~Savolainen, R.~S. Cowan, A.~P. Vogler, G.~K. Roderick, and R.~Lane.
\newblock Towards writing the encyclopaedia of life: an introduction to {DNA}
  barcoding.
\newblock {\em Phil. Trans. R. Soc. B}, 360:1805--1811, 2005.

\bibitem{r52}
J.~Rothstein.
\newblock Organization and entropy.
\newblock {\em J. Appl. Phys.}, 23:1281--1282, 1952.

\bibitem{w60}
S.~Watanabe.
\newblock Information theoretical analysis of multivariate correlation.
\newblock {\em IBM J. Res. Dev.}, 4:66--82, 1960.

\bibitem{cb15}
K.~K. Cassiano and V.~C. Barbosa.
\newblock Information integration in elementary cellular automata.
\newblock {\em J. Cell. Autom.}, 10:235--260, 2015.

\bibitem{b17}
V.~C. Barbosa.
\newblock Information integration from distributed threshold-based
  interactions.
\newblock {\em Complexity}, 2017:7046359, 2017.

\bibitem{bt08}
D.~Balduzzi and G.~Tononi.
\newblock Integrated information in discrete dynamical systems: motivation and
  theoretical framework.
\newblock {\em PLoS Comput. Biol.}, 4:e1000091, 2008.

\bibitem{nb11}
A.~Nathan and V.~C. Barbosa.
\newblock Network algorithmics and the emergence of information integration in
  cortical models.
\newblock {\em Phys. Rev. E}, 84:011904, 2011.

\bibitem{oat14}
M.~Oizumi, L.~Albantakis, and G.~Tononi.
\newblock From the phenomenology to the mechanisms of consciousness: integrated
  information theory 3.0.
\newblock {\em PLoS Comput. Biol.}, 10:e1003588, 2014.

\bibitem{aw10}
H.~Abdi and L.~J. Williams.
\newblock Principal component analysis.
\newblock {\em WIREs Comput. Stat.}, 2:433--459, 2010.

\bibitem{p03}
L.~Paninski.
\newblock Estimation of entropy and mutual information.
\newblock {\em Neural Comput.}, 15:1191--1253, 2003.

\bibitem{hs09}
J.~Hausser and K.~Strimmer.
\newblock Entropy inference and the {James}-{Stein} estimator, with application
  to nonlinear gene association networks.
\newblock {\em J. Mach. Learn. Res.}, 10:1469--1484, 2009.

\bibitem{app14}
E.~Archer, I.~M. Park, and J.~W. Pillow.
\newblock Bayesian entropy estimation for countable discrete distributions.
\newblock {\em J. Mach. Learn. Res.}, 15:2833--2868, 2014.

\bibitem{mtasw11}
C.~Mora, D.~P. Tittensor, S.~Adl, A.~G.~B. Simpson, and B.~Worm.
\newblock How many species are there on {E}arth and in the ocean?
\newblock {\em PLoS Biol.}, 9:e1001127, 2011.

\bibitem{sjpl12}
B.~R. Scheffers, L.~N. Joppa, S.~L. Pimm, and W.~F. Laurance.
\newblock What we know and don'’t know about {E}arth'’s missing
  biodiversity.
\newblock {\em Trends Ecol. Evol.}, 27:501--510, 2012.

\bibitem{lmrw17}
B.~R. Larsen, E.~C. Miller, M.~K. Rhodes, and J.~J. Wiens.
\newblock Inordinate fondness multiplied and redistributed: the number of
  species on {E}arth and the new pie of life.
\newblock {\em Q. Rev. Biol.}, 92:229--265, 2017.

\end{thebibliography}
\bibliographystyle{unsrt}

\newpage
\appendix
\renewcommand\thetable{\thesection.\arabic{table}}
\setcounter{table}{0}
\renewcommand\thefigure{\thesection.\arabic{figure}}
\setcounter{figure}{0}

\section{Supplementary material}
This appendix collects Table~\ref{nseqs} and Figures~\ref{ngtc-app}
and~\ref{scree}.

\begin{table}[p]
\scriptsize
\centering
\caption{Number of sequences downloaded (N.D.) and resulting from screening for
each value of $n$.}
\label{nseqs}
\begin{tabular}{lrrrrrr}
\hline
\multicolumn{1}{c}{Phylum}
&\multicolumn{1}{c}{N.D.}
&\multicolumn{1}{c}{$n=432$}
&\multicolumn{1}{c}{$n=459$}
&\multicolumn{1}{c}{$n=486$}
&\multicolumn{1}{c}{$n=513$}
&\multicolumn{1}{c}{$n=540$} \\
\hline
Acanthocephala   &         $465$ &         $341$ &         $243$ &         $243$ &         $242$ &         $239$ \\
Annelida         &     $31\,700$ &     $22\,931$ &     $22\,850$ &     $22\,658$ &     $22\,048$ &     $21\,581$ \\
Arthropoda       & $3\,854\,064$ & $3\,054\,118$ & $3\,040\,943$ & $3\,005\,920$ & $2\,987\,772$ & $2\,884\,779$ \\
Brachiopoda      &          $85$ &          $12$ &          $12$ &          $12$ &          $12$ &          $12$ \\
Bryozoa          &      $1\,611$ &      $1\,303$ &      $1\,126$ &      $1\,028$ &         $897$ &         $894$ \\
Chaetognatha     &         $287$ &         $185$ &         $185$ &         $132$ &         $132$ &         $131$ \\
Chordata         &    $304\,239$ &    $250\,169$ &    $249\,322$ &    $247\,814$ &    $246\,555$ &    $243\,226$ \\
Cnidaria         &      $9\,593$ &      $6\,667$ &      $6\,374$ &      $6\,181$ &      $6\,051$ &      $5\,824$ \\
Cycliophora      &         $273$ &          $31$ &          $31$ &          $31$ &           $0$ &           $0$ \\
Echinodermata    &     $19\,939$ &     $13\,080$ &     $13\,072$ &     $12\,890$ &     $12\,662$ &     $12\,549$ \\
Gnathostomulida  &           $8$ &           $0$ &           $0$ &           $0$ &           $0$ &           $0$ \\
Hemichordata     &          $12$ &          $10$ &          $10$ &          $10$ &          $10$ &          $10$ \\
Heterokontophyta &      $4\,366$ &      $4\,263$ &      $4\,207$ &      $4\,200$ &      $4\,059$ &      $4\,056$ \\
Mollusca         &     $99\,413$ &     $61\,792$ &     $60\,719$ &     $59\,988$ &     $58\,305$ &     $55\,758$ \\
Nematoda         &      $4\,894$ &      $3\,133$ &      $2\,637$ &      $2\,632$ &      $2\,625$ &      $2\,544$ \\
Nemertea         &         $942$ &         $613$ &         $612$ &         $612$ &         $607$ &         $593$ \\
Onychophora      &         $658$ &         $450$ &         $379$ &         $379$ &         $379$ &         $354$ \\
Platyhelminthes  &      $8\,297$ &      $5\,609$ &      $5\,346$ &      $3\,917$ &      $3\,837$ &      $3\,748$ \\
Porifera         &      $2\,133$ &      $1\,654$ &      $1\,638$ &      $1\,625$ &      $1\,326$ &      $1\,292$ \\
Priapulida       &           $4$ &           $3$ &           $3$ &           $3$ &           $3$ &           $3$ \\
Rhodophyta       &     $29\,703$ &     $21\,121$ &     $21\,082$ &     $20\,759$ &     $20\,730$ &     $20\,659$ \\
Rotifera         &      $6\,261$ &      $4\,995$ &      $4\,968$ &      $4\,899$ &      $3\,819$ &      $3\,742$ \\
Sipuncula        &         $363$ &         $299$ &         $298$ &         $298$ &         $298$ &         $298$ \\
Tardigrada       &         $729$ &         $700$ &         $694$ &         $692$ &         $653$ &         $605$ \\
Xenoturbellida   &           $2$ &           $1$ &           $1$ &           $1$ &           $1$ &           $1$ \\
\cline{2-7}
&\multicolumn{1}{c}{$n=567$}
&\multicolumn{1}{c}{$n=594$}
&\multicolumn{1}{c}{$n=621$}
&\multicolumn{1}{c}{$n=648$}
&\multicolumn{1}{c}{$n=675$}
&\multicolumn{1}{c}{$n=702$} \\
\cline{2-7}
Acanthocephala   &         $239$ &         $224$ &         $130$ &          $87$ &           $3$ &           $1$ \\
Annelida         &     $21\,083$ &     $20\,582$ &     $19\,551$ &     $17\,674$ &         $908$ &         $608$ \\
Arthropoda       & $2\,613\,568$ & $1\,860\,696$ & $1\,653\,239$ & $1\,432\,478$ &     $54\,934$ &     $44\,105$ \\
Brachiopoda      &           $7$ &           $7$ &           $6$ &           $5$ &           $1$ &           $1$ \\
Bryozoa          &         $815$ &         $479$ &         $419$ &         $379$ &         $234$ &          $12$ \\
Chaetognatha     &         $104$ &          $98$ &          $86$ &          $42$ &          $26$ &          $11$ \\
Chordata         &    $239\,945$ &    $234\,347$ &    $224\,732$ &    $203\,669$ &     $59\,286$ &     $39\,800$ \\
Cnidaria         &      $5\,698$ &      $5\,414$ &      $4\,859$ &      $4\,276$ &      $1\,835$ &      $1\,291$ \\
Cycliophora      &           $0$ &           $0$ &           $0$ &           $0$ &           $0$ &           $0$ \\
Echinodermata    &     $11\,986$ &     $11\,626$ &     $10\,971$ &      $9\,858$ &      $1\,618$ &      $1\,314$ \\
Gnathostomulida  &           $0$ &           $0$ &           $0$ &           $0$ &           $0$ &           $0$ \\
Hemichordata     &          $10$ &          $10$ &          $10$ &           $9$ &           $6$ &           $6$ \\
Heterokontophyta &      $3\,827$ &      $3\,798$ &      $3\,763$ &      $3\,731$ &      $3\,553$ &      $2\,380$ \\
Mollusca         &     $53\,196$ &     $49\,526$ &     $44\,305$ &     $33\,148$ &      $5\,610$ &      $3\,754$ \\
Nematoda         &      $2\,411$ &      $2\,314$ &      $2\,127$ &      $1\,513$ &         $556$ &         $471$ \\
Nemertea         &         $548$ &         $485$ &         $452$ &         $395$ &          $49$ &          $25$ \\
Onychophora      &         $353$ &         $227$ &         $168$ &          $56$ &          $53$ &          $52$ \\
Platyhelminthes  &      $3\,658$ &      $3\,416$ &      $3\,102$ &      $2\,937$ &      $2\,784$ &      $2\,610$ \\
Porifera         &      $1\,192$ &         $832$ &         $774$ &         $618$ &         $259$ &         $153$ \\
Priapulida       &           $3$ &           $3$ &           $3$ &           $3$ &           $1$ &           $1$ \\
Rhodophyta       &     $20\,558$ &     $20\,407$ &     $20\,193$ &     $19\,668$ &      $4\,865$ &      $4\,543$ \\
Rotifera         &      $3\,606$ &      $3\,119$ &      $2\,594$ &      $1\,796$ &         $234$ &         $219$ \\
Sipuncula        &         $289$ &         $282$ &         $213$ &         $195$ &          $49$ &          $48$ \\
Tardigrada       &         $553$ &         $548$ &         $522$ &         $406$ &         $175$ &          $58$ \\
Xenoturbellida   &           $1$ &           $1$ &           $1$ &           $1$ &           $1$ &           $1$ \\
\hline
\end{tabular}

\end{table}

\begin{figure}[p]
\centering
\includegraphics[scale=0.80]{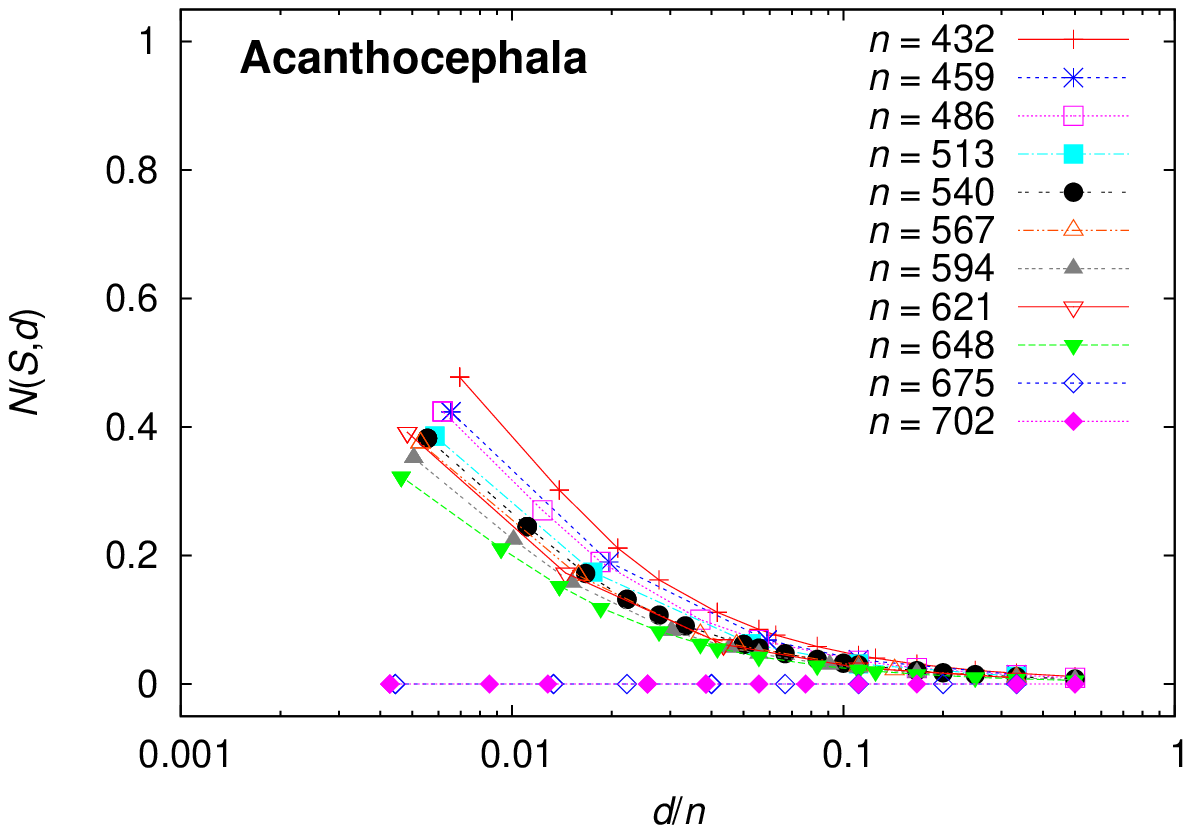}\\
\includegraphics[scale=0.80]{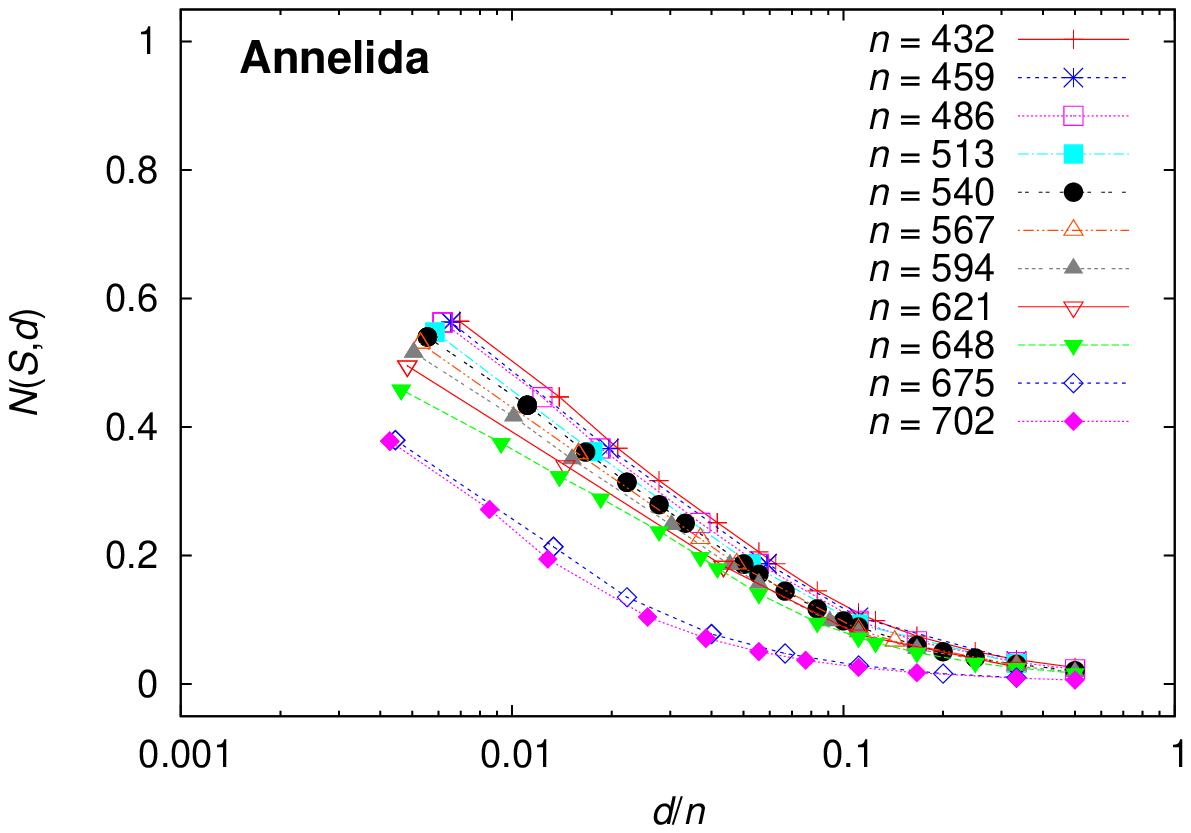}
\caption{The $107$ $N(S,d)$ descriptors for each of the $23$ phyla studied.}
\label{ngtc-app}
\end{figure}

\addtocounter{figure}{-1}
\begin{figure}[p]
\centering
\includegraphics[scale=0.80]{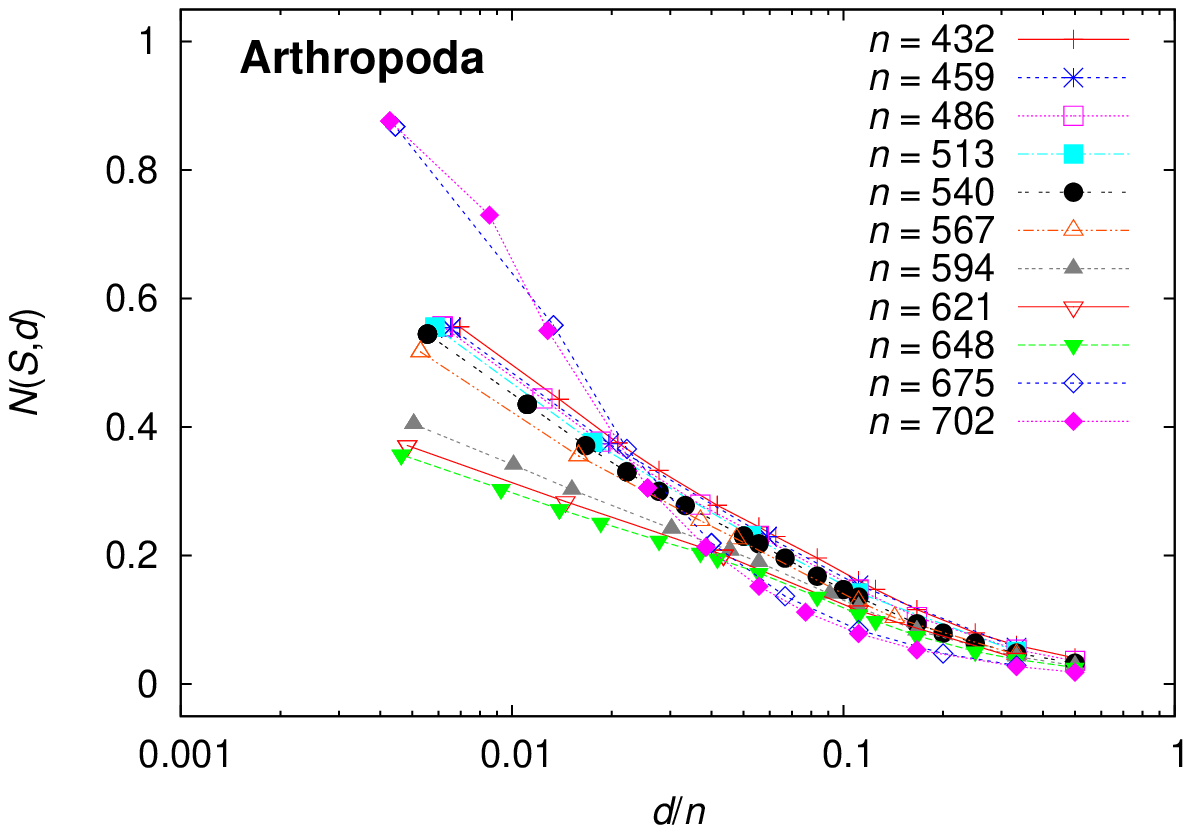}\\
\includegraphics[scale=0.80]{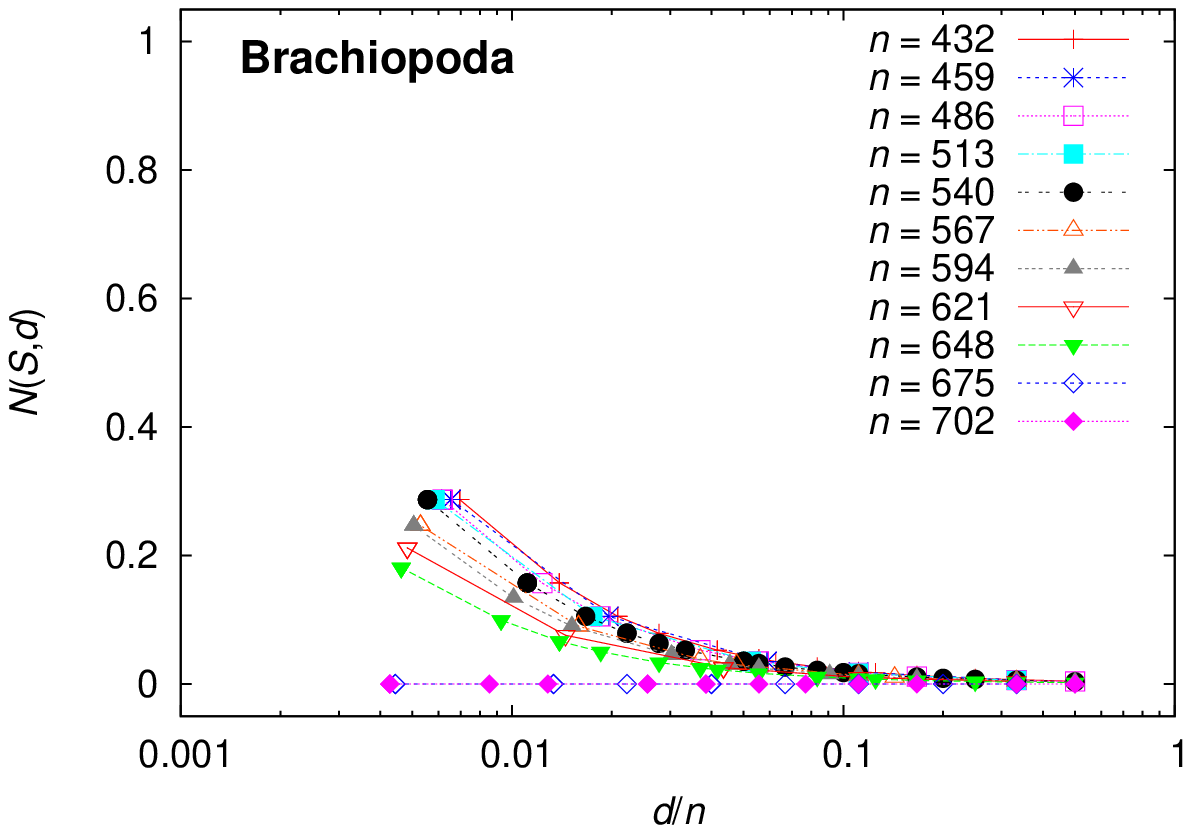}
\caption{Continued.}
\end{figure}

\addtocounter{figure}{-1}
\begin{figure}[p]
\centering
\includegraphics[scale=0.80]{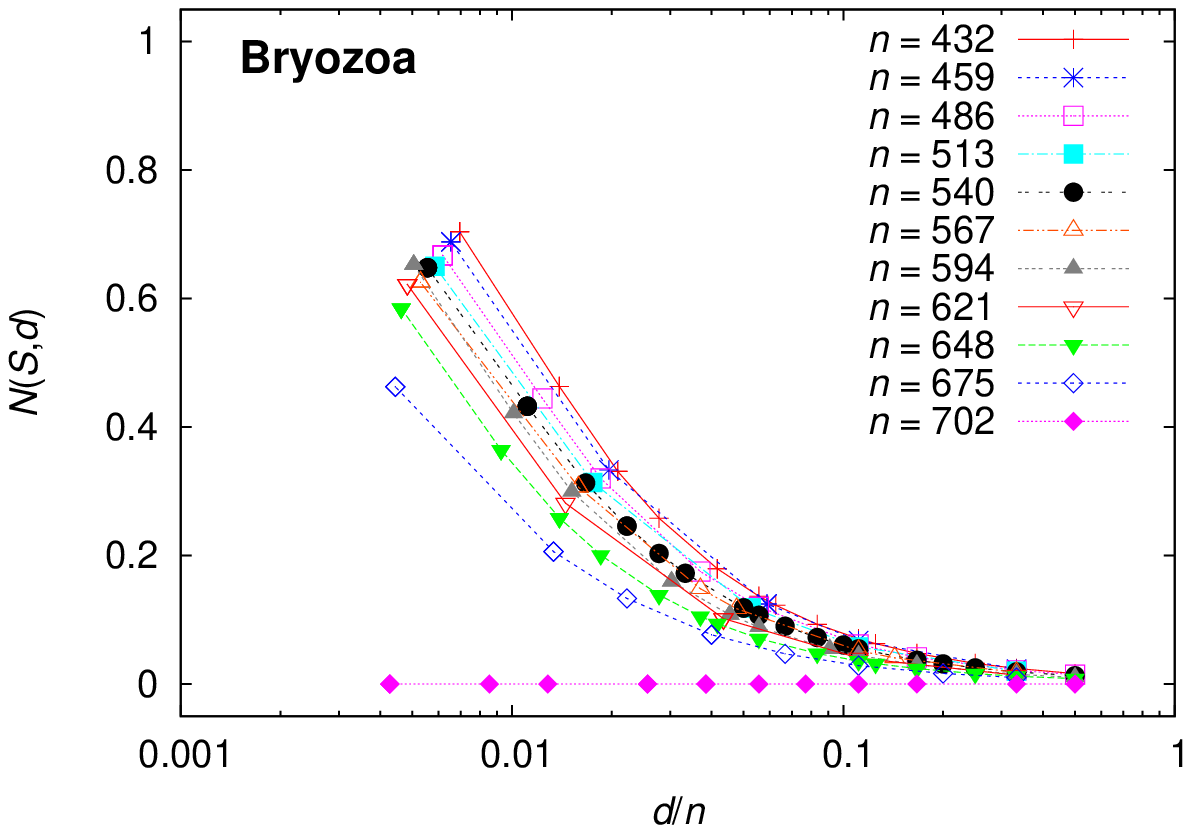}\\
\includegraphics[scale=0.80]{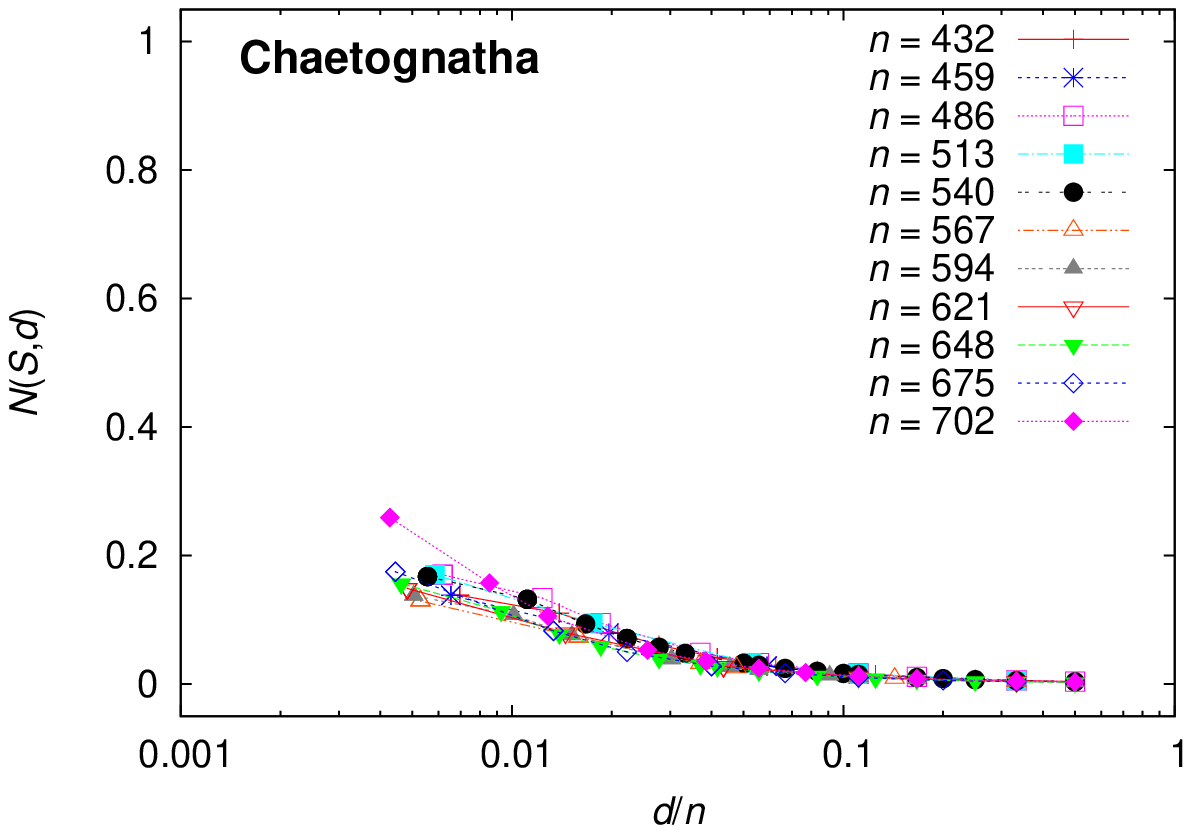}
\caption{Continued.}
\end{figure}

\addtocounter{figure}{-1}
\begin{figure}[p]
\centering
\includegraphics[scale=0.80]{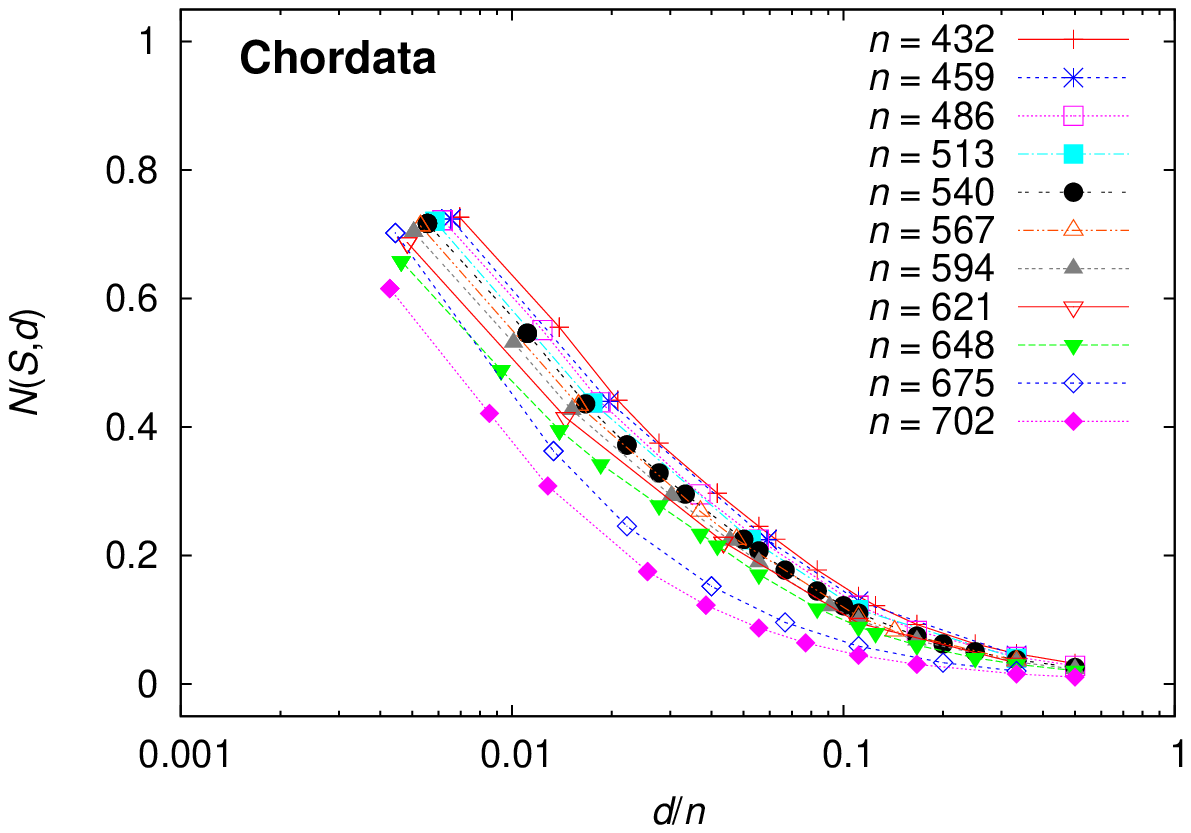}\\
\includegraphics[scale=0.80]{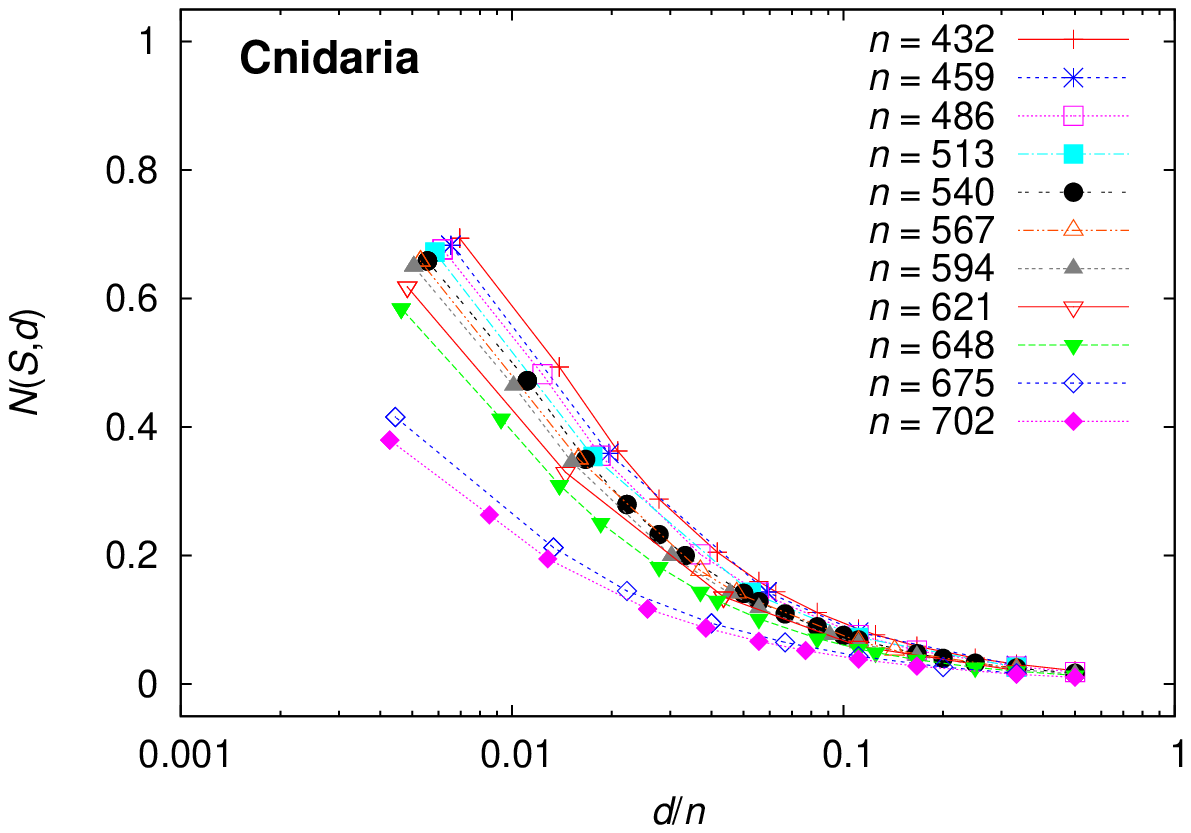}
\caption{Continued.}
\end{figure}

\addtocounter{figure}{-1}
\begin{figure}[p]
\centering
\includegraphics[scale=0.80]{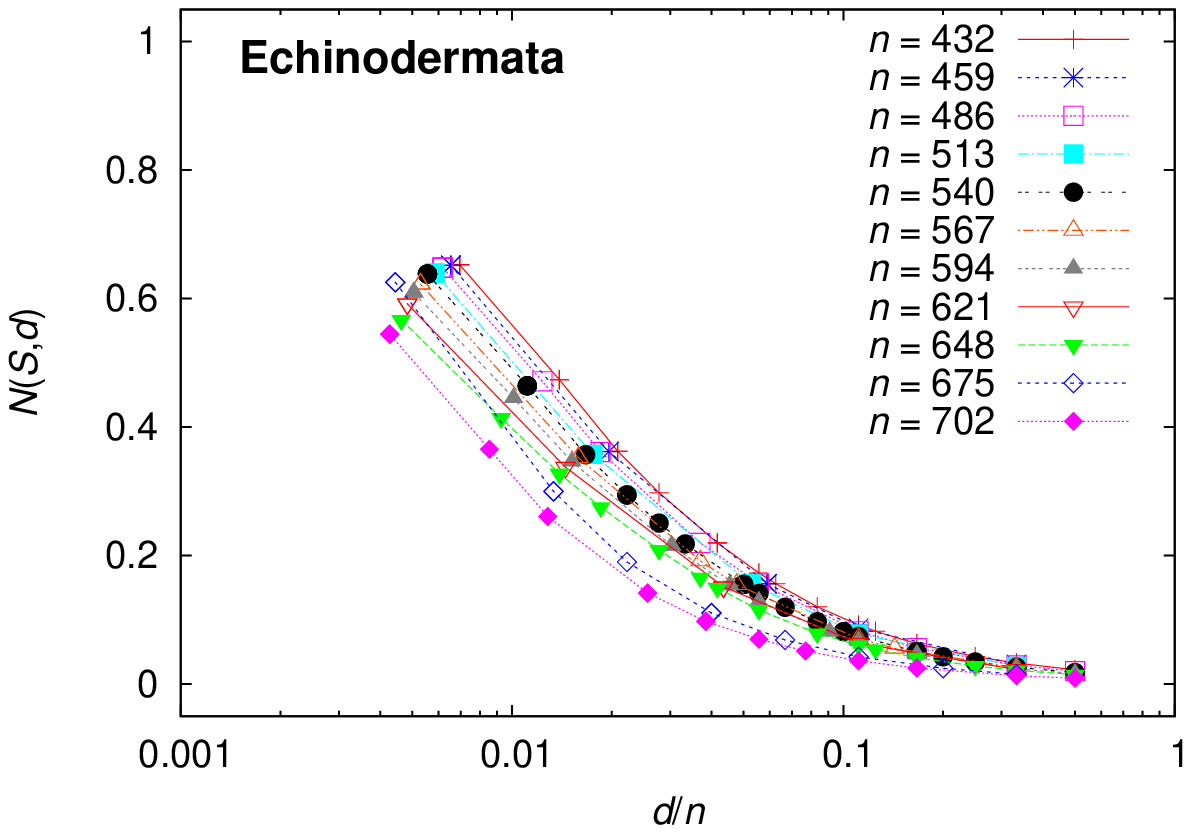}\\
\includegraphics[scale=0.80]{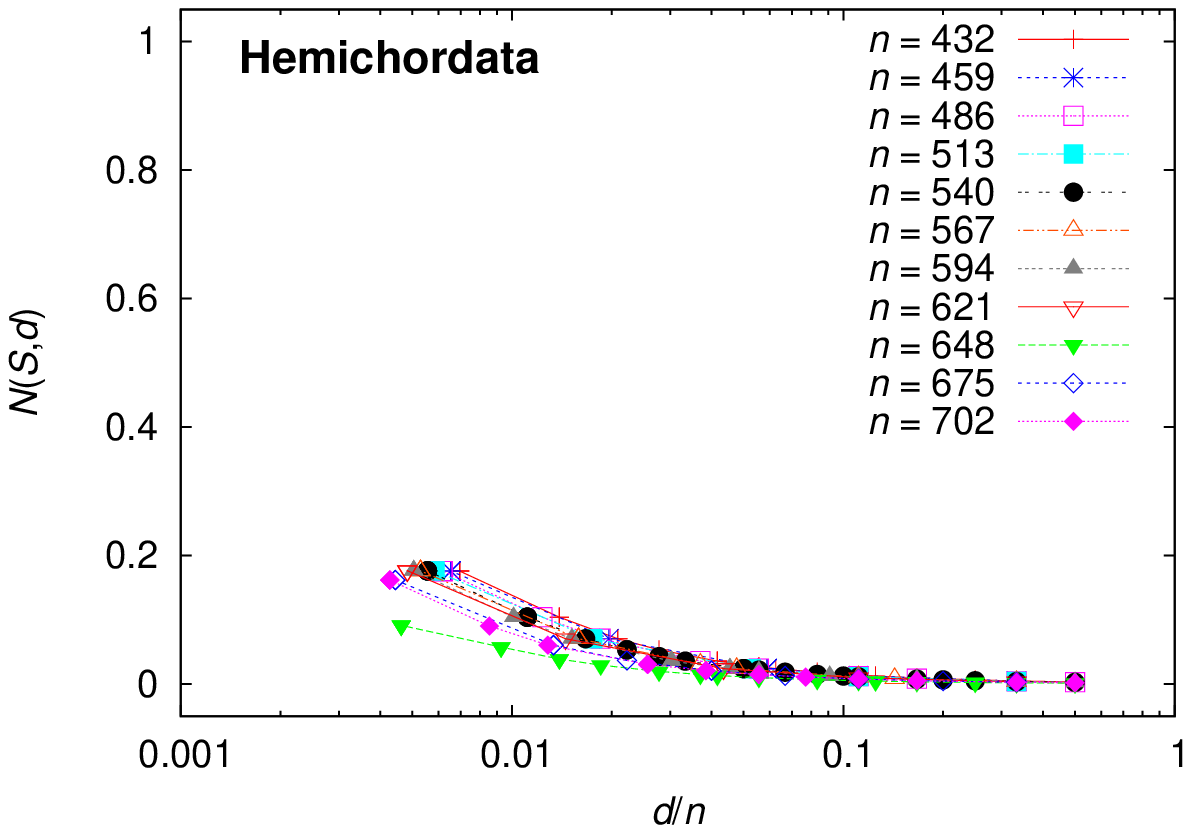}
\caption{Continued.}
\end{figure}

\addtocounter{figure}{-1}
\begin{figure}[p]
\centering
\includegraphics[scale=0.80]{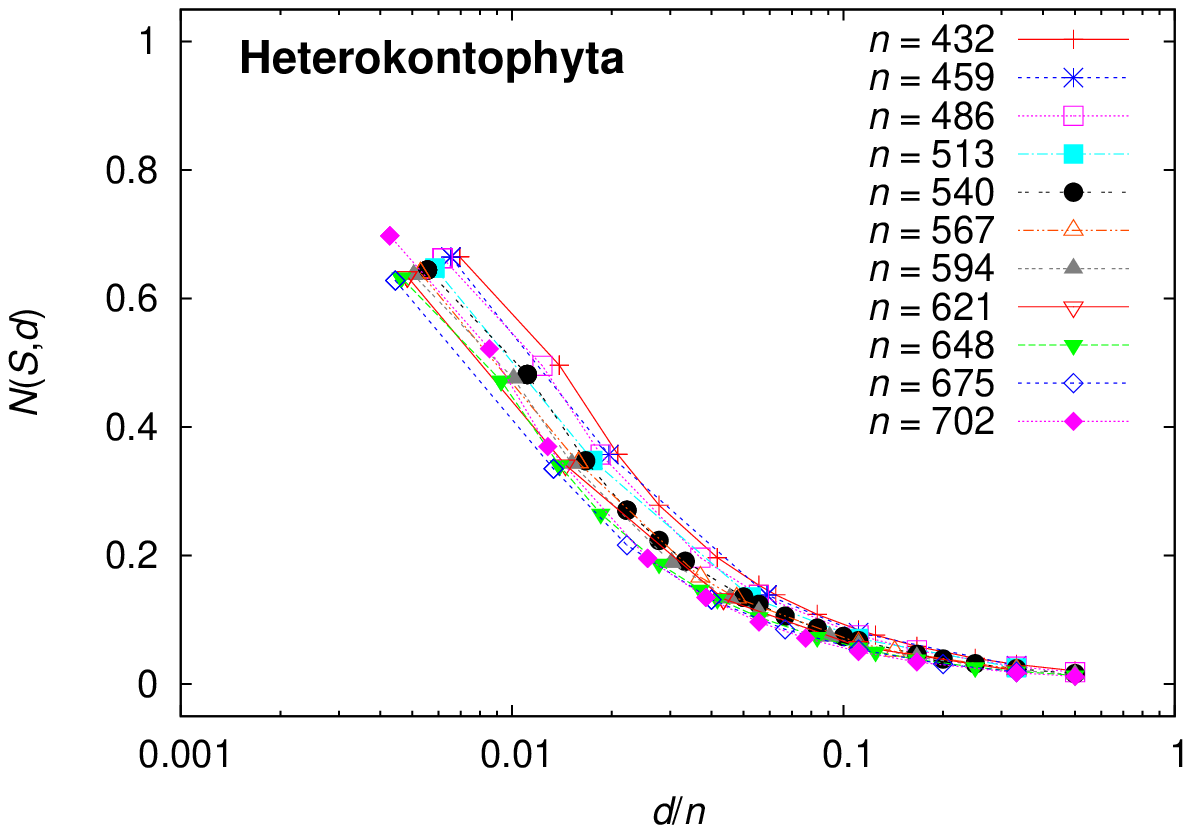}\\
\includegraphics[scale=0.80]{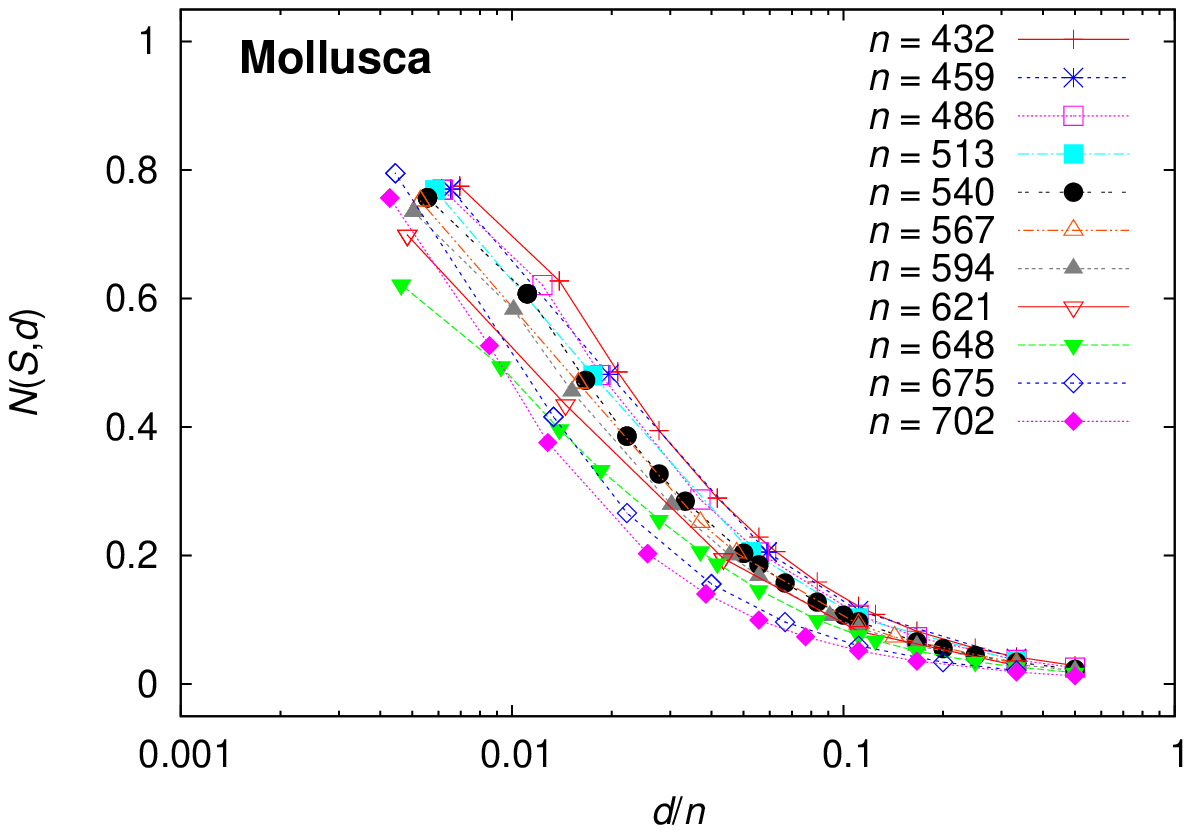}
\caption{Continued.}
\end{figure}

\addtocounter{figure}{-1}
\begin{figure}[p]
\centering
\includegraphics[scale=0.80]{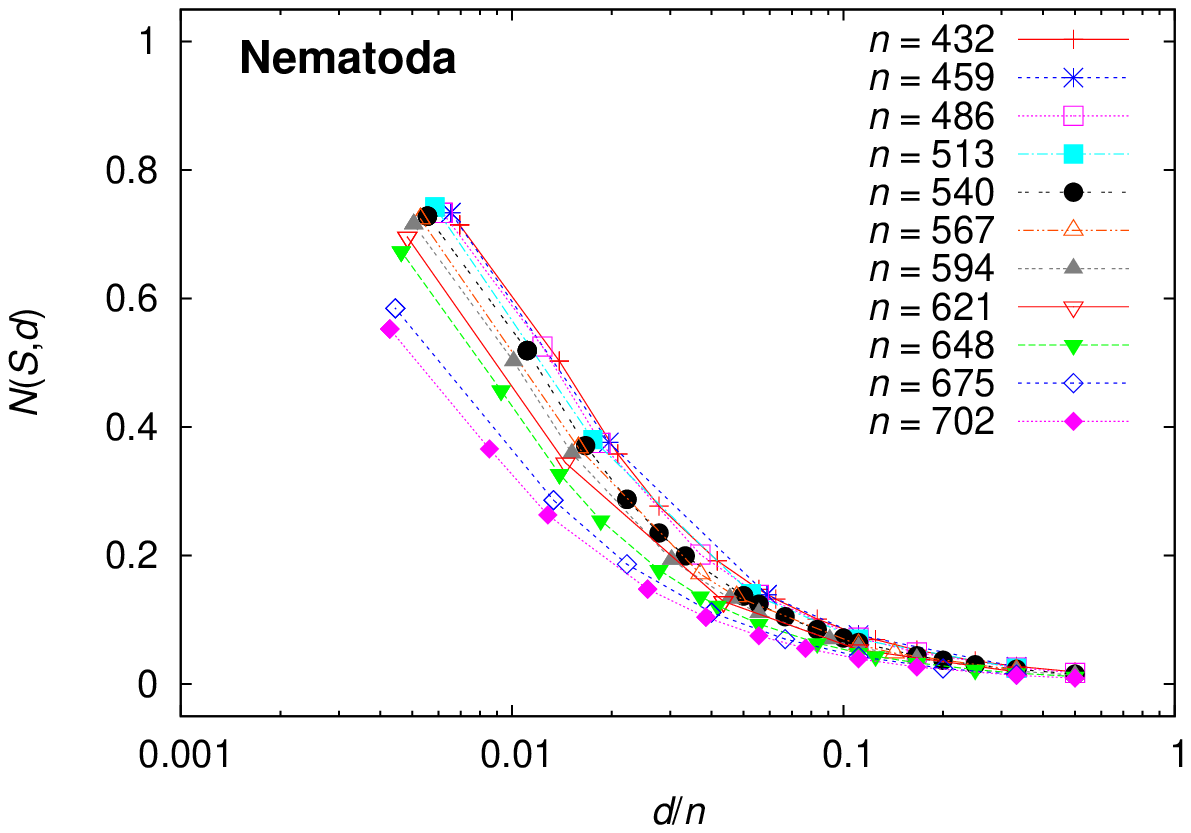}\\
\includegraphics[scale=0.80]{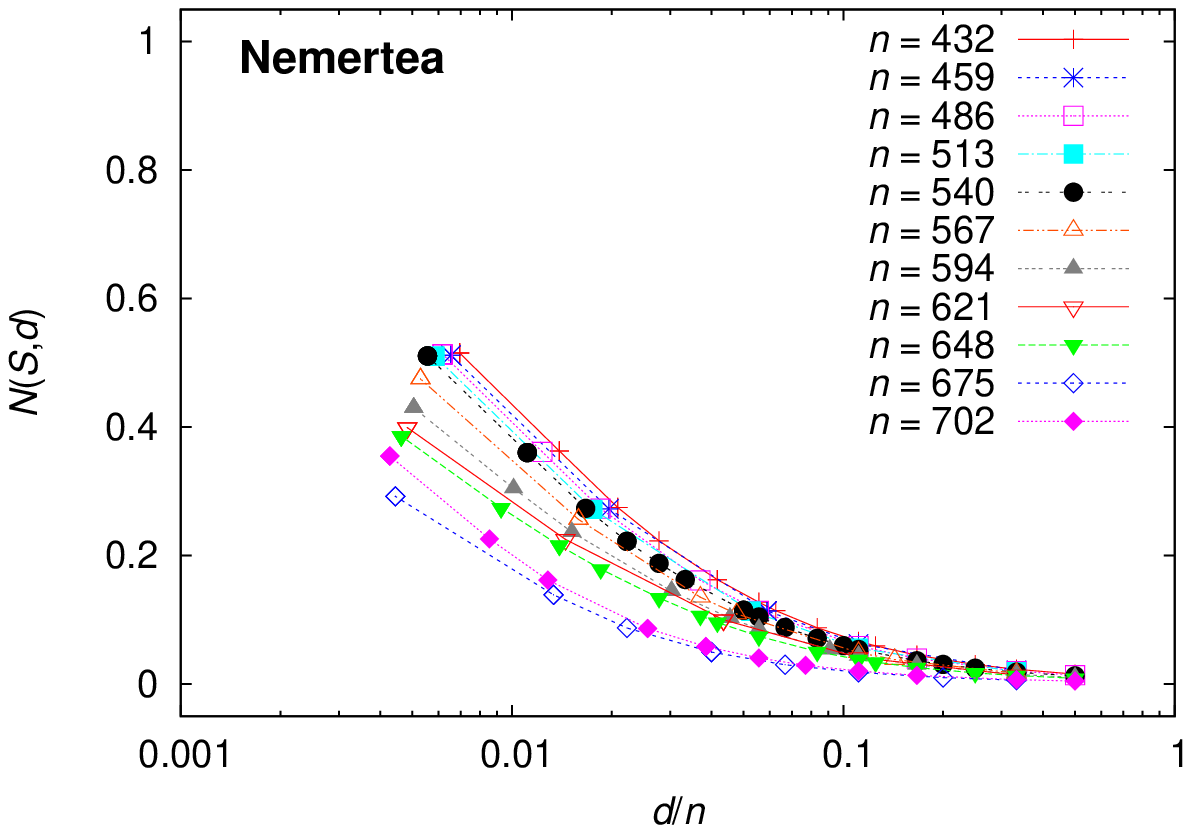}
\caption{Continued.}
\end{figure}

\addtocounter{figure}{-1}
\begin{figure}[p]
\centering
\includegraphics[scale=0.80]{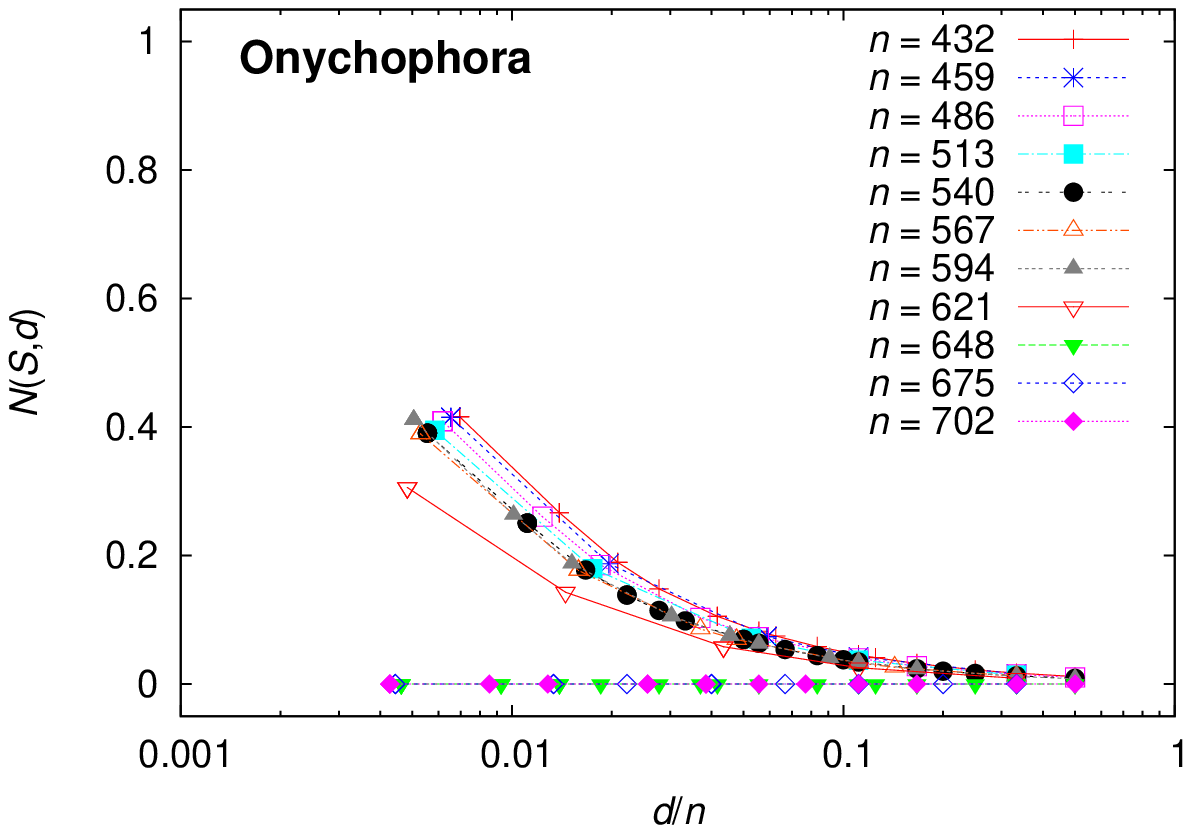}\\
\includegraphics[scale=0.80]{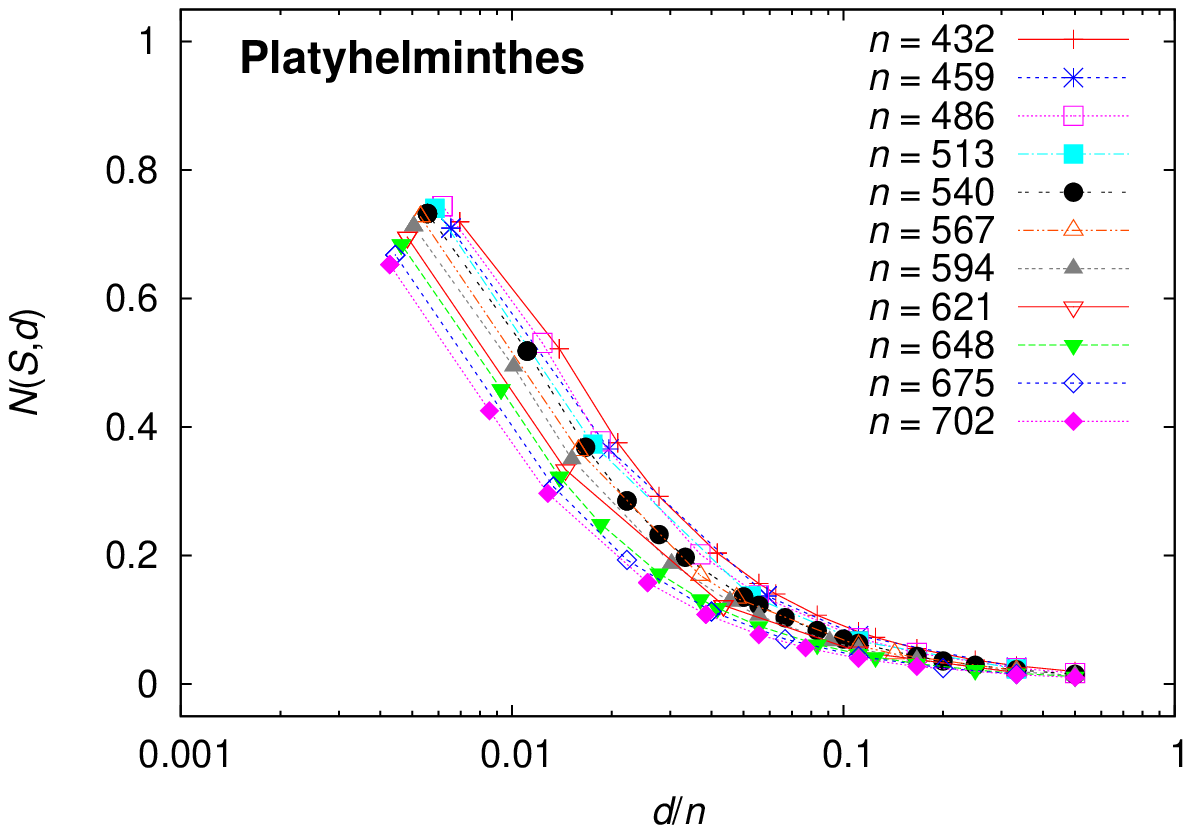}
\caption{Continued.}
\end{figure}

\addtocounter{figure}{-1}
\begin{figure}[p]
\centering
\includegraphics[scale=0.80]{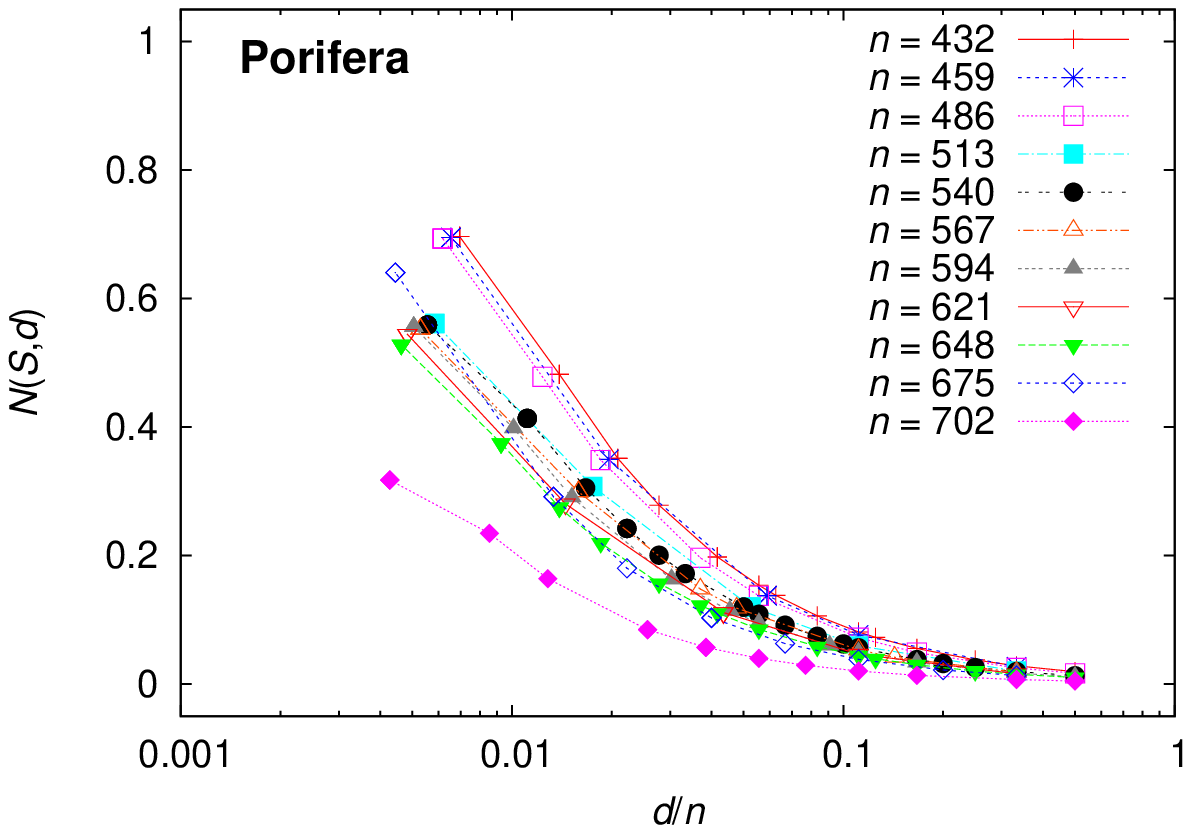}\\
\includegraphics[scale=0.80]{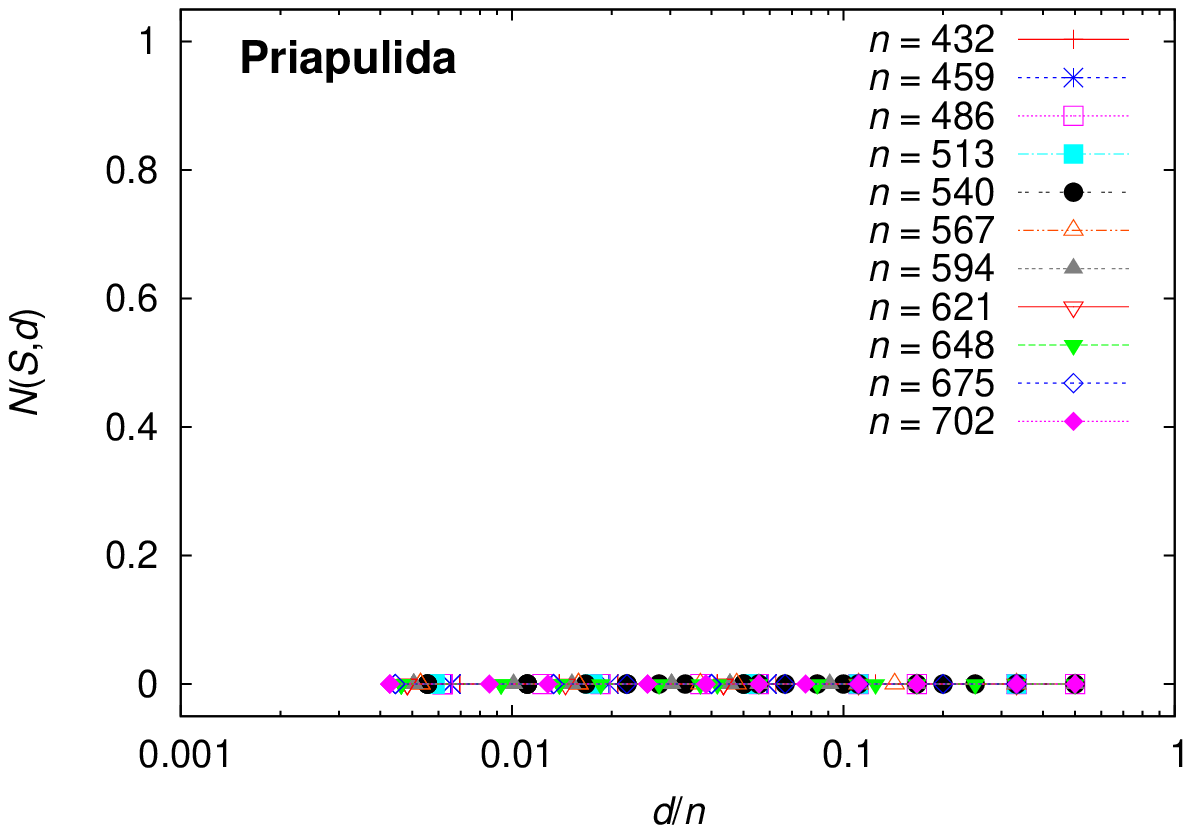}
\caption{Continued.}
\end{figure}

\addtocounter{figure}{-1}
\begin{figure}[p]
\centering
\includegraphics[scale=0.80]{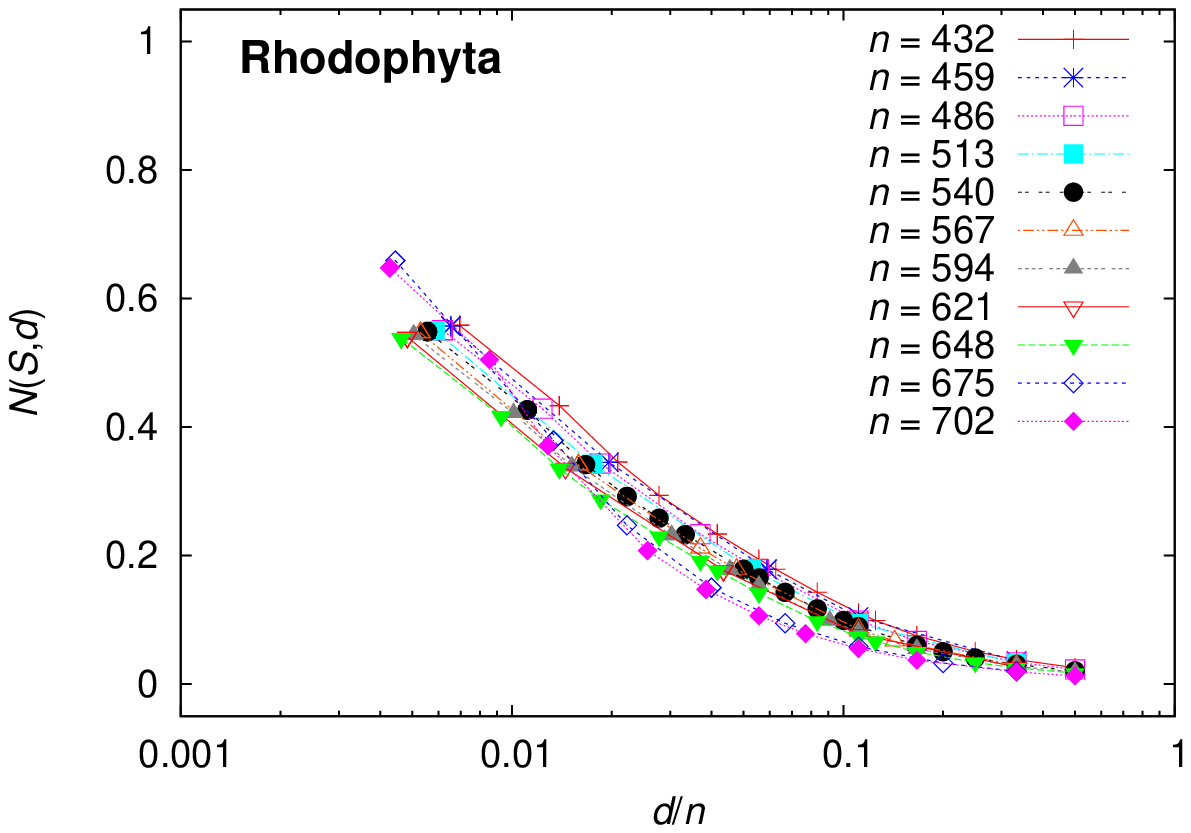}\\
\includegraphics[scale=0.80]{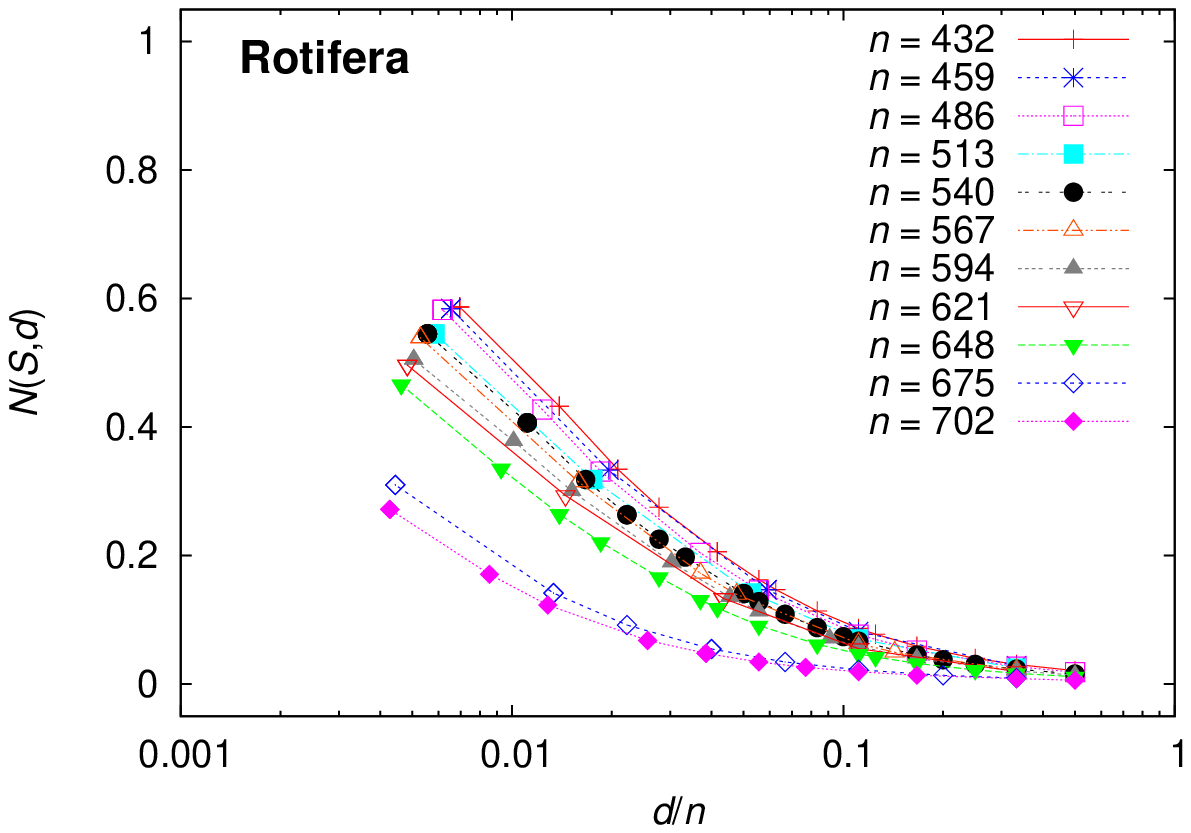}
\caption{Continued.}
\end{figure}

\addtocounter{figure}{-1}
\begin{figure}[p]
\centering
\includegraphics[scale=0.80]{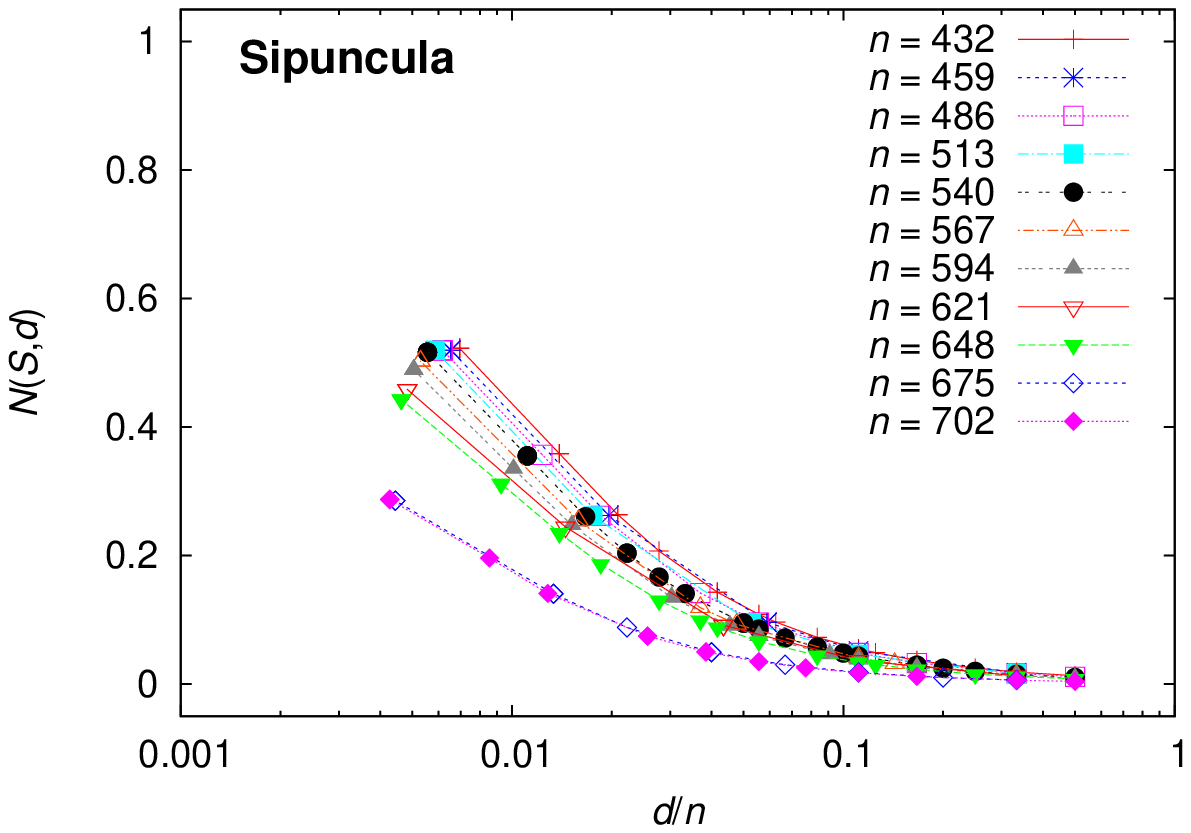}\\
\includegraphics[scale=0.80]{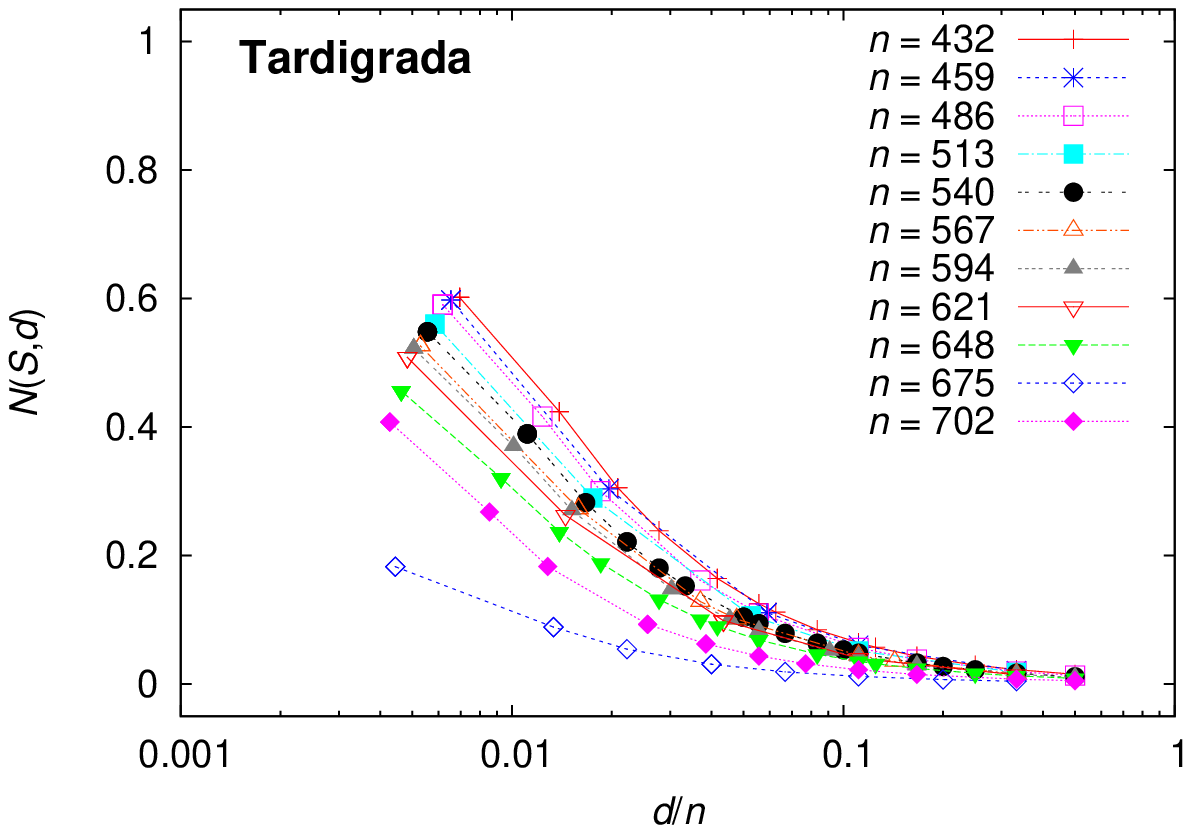}
\caption{Continued.}
\end{figure}

\addtocounter{figure}{-1}
\begin{figure}[p]
\centering
\includegraphics[scale=0.80]{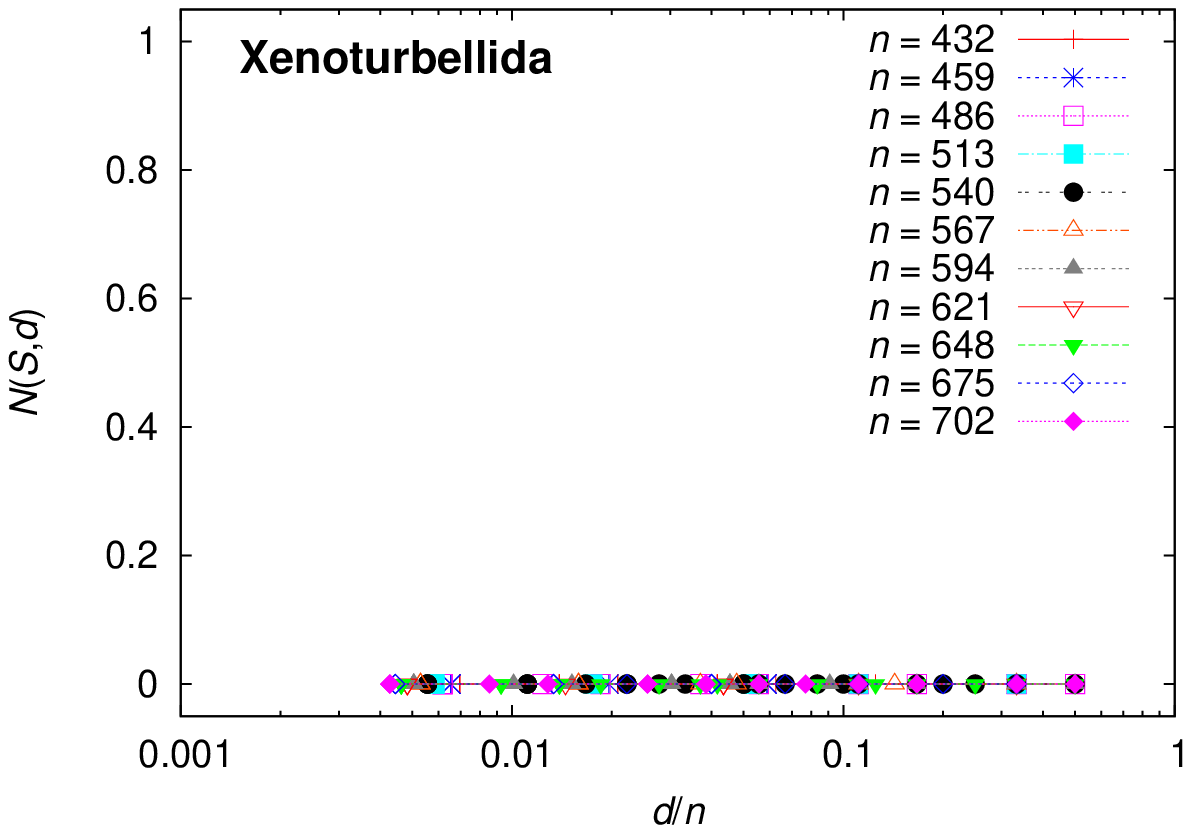}\\
\caption{Continued.}
\end{figure}

\begin{figure}[p]
\centering
\includegraphics[scale=0.80]{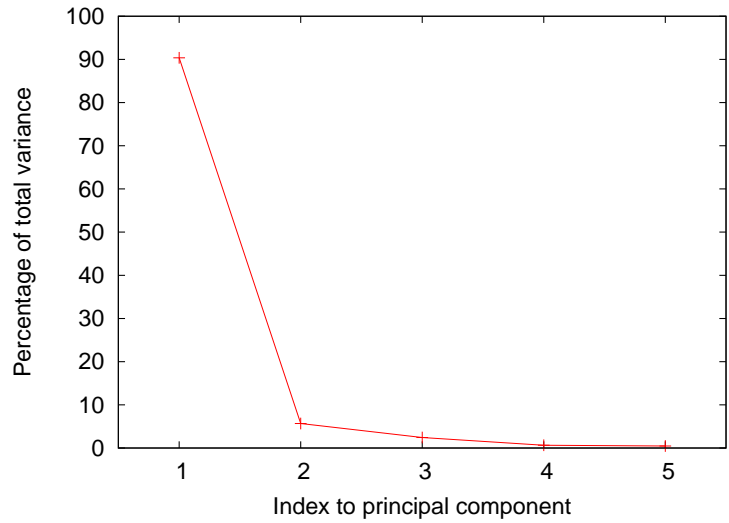}\\
\caption{Scree plot resulting from PCA on the $107$ standardized $N(S,d)$
descriptors. Only the first five principal components are included.}
\label{scree}
\end{figure}

\end{document}